\documentclass[final,5p,times,twocolumn,authoryear]{elsarticle}

\usepackage{amssymb}
\usepackage{amsmath,upgreek}
\usepackage{multicol,caption}
\usepackage{float,placeins}
\usepackage[T1]{fontenc}
\usepackage[colorlinks]{hyperref}

\usepackage{lineno}
\usepackage{hyperref}

\usepackage{comment}
\usepackage{multirow}

\journal{New Astronomy}

\begin{document}

\begin{frontmatter}



\title{Structural Analysis of Brightest Cluster Galaxies in Poor and Rich Clusters}


\author[SP1,EG]{Eman Shaaban}
\author[IU]{Sinan Alis\corref{mycorrespondingauthor}}
\cortext[mycorrespondingauthor]{Corresponding author}
\ead{salis@istanbul.edu.tr}
\author[SP]{Mehmet Bektasoglu}
\author[IU]{F. Korhan Yelkenci}
\author[HUA,IU2]{E. Kaan Ulgen}
\author[MQ,ARC]{Oguzhan Cakir}
\author[IU]{Suleyman Fisek}

\address[SP1]{Department of Physics, Sakarya University, Sakarya, Turkey}
\address[EG]{National Research Center (NRC), Department of Physics, Giza, Cairo, Egypt}
\address[IU]{Department of Astronomy and Space Sciences, Faculty of Science, Istanbul University, 34116 Istanbul, Turkey}
\address[SP]{Department of Physics, Sakarya University, Sakarya, Turkey}
\address[IU2]{Astronomy and Space Sciences Program, Institute of Graduate Studies in Sciences, Istanbul University, 34116 Istanbul, Turkey}
\address[HUA]{Huawei Turkey, R\&D Center, Istanbul, Turkey}
\address[MQ]{School of Mathematical and Physical Sciences, Macquarie University, NSW 2109, Australia}
\address[ARC]{ARC Centre of Excellence for All Sky Astrophysics in 3 Dimensions (ASTRO 3D), Australia}

\begin{abstract}
Studying structural parameters of brightest cluster galaxies (BCGs) provides important clues to understand their formation and evolution. We present the results of the surface brightness profile fitting of 1685 brightest cluster galaxies (BCGs) drawn from the Canada-France-Hawaii Telescope Legacy Survey in the redshift range of $0.1 < z < 1.0$. We fit $r$-band images of BCGs with a single Sérsic profile. The sample is splitted into two groups based on the host cluster richness to investigate the impact of the environment. Our results suggest that BCGs in rich clusters are statistically larger than their counterparts in poor clusters. We provide best-fit linear regressions for the Kormendy, the $log \ R_e - log \ n$, and the size-luminosity relations. In addition, we examined the evolution of the structural parameters, however the BCGs in our sample do not show a significant size change since z$\sim$1.
\end{abstract}


\end{frontmatter}


\section{Introduction}

Observations and simulations indicate that the core of galaxy clusters are dominated by the most massive and the most luminous galaxies in the Universe known as brightest cluster galaxies (BCGs). These galaxies show unique properties in their sizes, dark matter contents and the velocity dispersions compared to the normal elliptical galaxies \citep{VonDerLinden2007}. 

Low star formation rates of BCGs imply mass-growth via dry-mergers \citep{Liu2009,Ruszkowski2009,Lidman2012,Lavoie2016} which is consistent with the low-scatter in their luminosity due to the dissipationless processes \citep{PostmanLauer1995,AragonSalamanca1998,VonDerLinden2007}. As the BCG continues to grow through merging with the  surrounding satellite galaxies, its size, luminosity, and stellar mass, as well as the magnitude difference with  respect to other nearby cluster members (i.e. dominance), increase \citep{Bernardi2007}. However, \cite{Scarlata2007} and \cite{Lidman2013} suggest that major mergers, including gas-rich ones, should take place in the formation of BCGs in the similar redshift ranges at $z<1$.

Hierarchical formation of clusters anticipate a strong connection between the cluster halo and its BCG. In this scenario, the stellar mass of the BCGs is closely related to the mass of the dark matter halo in which it is formed. \cite{DeLucia2007} showed that the stellar masses of BCGs are assembled around z=0.5 with an evolutionary path consistent with the hierarchical growth of structures as suggested by the $\Lambda$CDM cosmology. Thus, investigating structural properties of BCGs play an important role in the understanding of their formation, and especially their evolution.

It has been confirmed that there is a strong correlation between BCG parameters and the main properties of their host clusters \citep{Nelson2002,Ascaso2011,Lidman2012}. That connection with their host clusters (e.g. environment) have been extensively studied by means of stellar mass, size, surface brightness profiles, and the merging events \citep{Brough2005,Stott2008,Brough2008,Hansen2009,Bai2014,Bellstedt2016}.

Position angle, as a structural parameter of a galaxy, can be used to investigate the alignment of BCGs with their host clusters \citep{Dubinski1998}. Such studies in low \citep{Fasano2010}, intermediate \citep{Niederste-Ostholt2010,Chu2022}, and high redshifts \citep{West2017} showed that BCGs are in general well aligned with their host clusters.

Although BCGs are predominantly elliptical in morphology, a large fraction of them exhibit an extended, low-surface-brightness envelope around the central region \citep{Dressler1984,Lauer1992,Oegerle2001,Zhao2015} which makes their brightness profiles different from those of regular ellipticals. The S\'ersic index ($n$) measures the concentration of the light profile in galaxies and can be used to quantify its structure \citep{Peng2002,Peng2010}. Sizes that obtained from these light profiles have been used to construct size-luminosity relations and it has been shown that it is different for BCGs than other early-type galaxies \citep{Bernardi2007,Samir2020}. Moreover, \cite{Tortorelli2018} examined member galaxies of the two intermediate-redshift clusters in the Hubble Frontier Fields and compared the Kormendy relations of early-type galaxies. They found similar slopes for the Kormendy relations for galaxies classified as early-type by S\'ersic index, as elliptical by visual inspection and as passive by spectral properties (e.g. star formation).

Size evolution of BCGs are quite controversial. There have been studies showing little or no evolution \citep{Stott2008,Chu2022}, and significant change in the galaxy size \citep{Bernardi2009,Ascaso2011}. This controversy when coupled with the assembly time of the BCGs becomes more important. Samples from both observations and simulations point out to an older stellar population which assembled relatively recent epochs (e.g. z$\sim$0.5). 

Homogeneity of the BCG properties \citep{Bernardi2007} makes them attractive for using as standard candles, which then can be used on the cosmological scales as they are the most luminous galaxies in the Universe.

In this study, we aim to investigate the impact of environment (via host cluster richness) on the BCG properties in a wide range of redshift (i.e. $0.1 < z < 1.0$). Using $r$-band images taken from CFHTLS survey, we apply S\'ersic profile fitting to the surface brightness of our sample galaxies. We examine the evolution of the structural parameters and the scaling relations for BCGs in different environments.

The paper is organized as follows: in Section 2, we present our data and the sample of BCGs. In Section 3, we describe our approach and the procedure. In Section 4 we present the results with discussion and we give our summary and conclusions in Section 5. Throughout the paper we use H$_{0}$ = 70 km s$^{-1}$ Mpc$^{-1}$, $\Omega_{ \textrm m}$ = 0.3 and $\Omega_{\Lambda}$ = 0.7.

\section{Data and Sample Selection}

\subsection{CFHTLS}
\label{sec:command}
The Canada-France-Hawaii Telescope Legacy Survey (CFHTLS) is an imaging survey carried out between 2003-2009 in the four wide and the four deep fields. The wide survey covers a total area of 171 deg$^{2}$ with imaging in $u^{*}g^{\prime}r^{\prime}i^{\prime}z^{\prime}$ bands. Due to the overlaps between adjacent fields, the effective area of the CFHTLS-Wide is 155 deg$^{2}$. Except the $u$-band, CFHTLS filters are almost identical with those used in the SDSS, hence we refer these filters as $ugriz$ for the rest of the paper. 

CFHTLS made use of MegaCam which is a mosaic CCD camera consisting of 36 thinned EEV detectors with each one has 2048 $\times$ 4612 pixels. This configuration has approximately 1 deg$\times$1 deg field of view for each MegaCam pointing. Processed individual images consisting of 19354$\times$19354 pixels with a pixel scale of 0.186 $^{\prime\prime}/ \textrm{pixel}$.
\begin{figure}[t]
\begin{center}
	\includegraphics[width=\columnwidth]{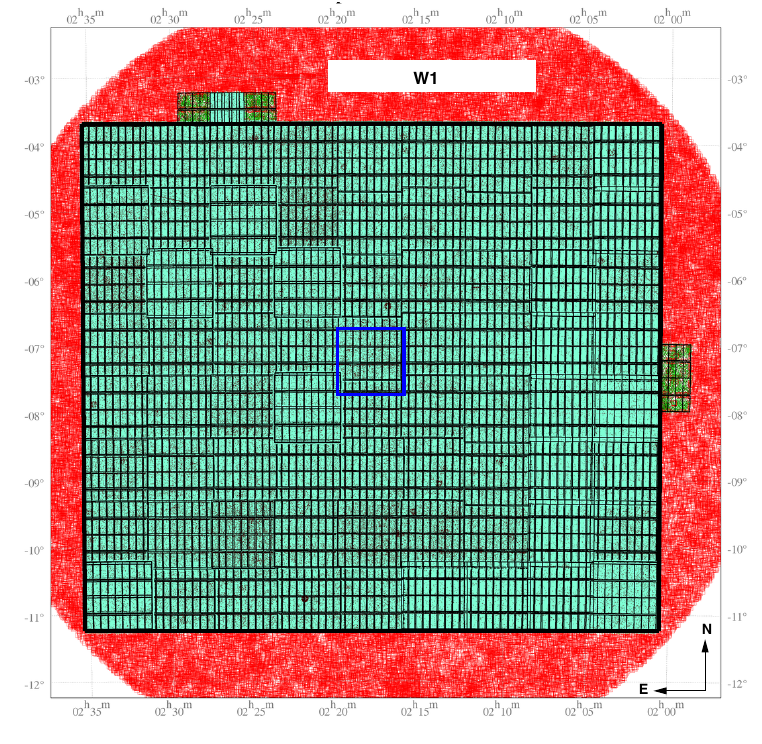}
	\caption{Geometry of CFHTLS W1 pointings. There are 9$\times$8 individual MegaCam pointings for this field. Solid black line represents the released image area with T0007 release. Regions outside the black line have been used for the astrometric calibrations. Due to overlaps between different MegaCam pointings total effective area covered with W1 is 63.75 deg$^{2}$ \citep{Hudelot12}. The blue rectangle at the center represents the reference field with central coordinates of RA=02$^h$ 18$^m$, Dec=-07$^\circ$ 00$^\prime$.}
    \label{fig:W1geom}
\end{center}
\end{figure}

In this study, we investigated the structural evolution of BCGs in clusters detected in the W1 field of the CFHTLS. The geometry and coordinates of the W1 field is given in Fig. \ref{fig:W1geom} which has a 9$\times$8 individual pointings. Since there are overlaps between the pointings, the total effective survey area in W1 is 63.75 deg$^{2}$. The median seeing of $r$-band images is 0.71$^{\prime\prime}$ for the W1 field where the $80\%$ completeness limit in $r$-band is 24 mag for extended sources.

Galaxy clusters, hence their BCGs, were detected from the W1 galaxy catalog where the regions around bright stars, ghosts, spikes, and other areas with lower cosmetic quality are masked. Thus, our object catalog contains 2,871,455 ($r\leq 24$) galaxies.

Data products of CFHTLS are images, mask files, object and photometric redshift catalogs. These products are processed, generated, and distributed by TERAPIX\footnote{A former data processing center at the Institut d'Astrophysique de Paris.}. Since the latest data release that took place in 2012 \citep{Hudelot12}, all data is publicly available and can be accessed via Canadian Astronomy Data Centre (CADC\footnote{https://www.cadc-ccda.hia-iha.nrc-cnrc.gc.ca/en/cfht/cfhtls.html}). 

The image processing pipeline of the CFHTLS was optimized for better object detection with accurate photometry rather than detecting low surface brightness features. The subtraction of sky background was likely removed such features from the images which could have impact on our effective radius determination. Thus, for lower redshifts (e.g. z<0.3) the surface brightness profile of BCGs might not be accurate enough based on the CFHTLS images. A detailed discussion on the importance of background subtraction can be found in \cite{Furnell2021}.

\subsection{Accuracy of Photometric Redshifts}

Photometric redshifts ($z_{p}$) of the CFHTLS have been computed by TERAPIX using \texttt{LePhare}. \texttt{LePhare} is a SED fitting tool that finds the best match with the compared template spectra by means of $\chi ^{2}$ minimization \citep{Ilbert06}. Five different template spectra (E, Sbc, Scd, Irr, and SB) were used for the photometric redshift computation for the CFHTLS fields \citep{Coupon2009}. These five template are extrapolated into 66 templates in order to cover redshift ranges up to z $\sim$ 1.2. Calibration of the templates have been done using spectroscopic redshifts ($z_{s}$) obtained by VVDS \citep{LeFevre05}.
\begin{table}[t]
\centering
\begin{tabular}{ c|c|c }
\hline
$i_{AB}$  & $\sigma_{\Delta z /(1+z_s)}$ & $\eta \ (\%)$ \\
\hline
20.5 & 0.025 & 1.12 \\ 
21.0 & 0.026 & 1.57  \\
21.5 & 0.029 & 1.39  \\
22.0 & 0.032 & 2.25  \\
22.5 & 0.037 & 2.81  \\
23.0 & 0.043 & 4.91  \\
23.5 & 0.048 & 7.63  \\
24.0 & 0.053 & 10.13 \\
\hline
\end{tabular}
\caption{Photometric redshift accuracy and outlier fraction for different magnitude cuts in i-band \citep{Coupon2009}.}
\label{tabzphot}
\end{table}

Comparison of photometric redshifts computed for CFHTLS with spectroscopic redshifts available for the W1 field reveals a mean error of 0.03 \citep{Coupon2009}. Table \ref{tabzphot} lists photometric redshift accuracy for different magnitude limits for CFHTLS-W1. The outlier fraction given in the table computed as the ratio of galaxies with $ \lvert \Delta z \rvert \geq 0.15 \times (1+z_{s})$ where $\Delta z$ is the difference between $z_{s}$ and $z_{p}$.

\subsection{BCG Sample}

BCG sample of the present study is derived from the galaxy clusters detected in CFHTLS-W1. Clusters in the W1 region were determined by the Wavelet Z-Photometric (\texttt{WaZP}) cluster finder. The (\texttt{WaZP}) cluster finder is prepared to discover galaxy clusters from multi-wavelength optical imaging galaxy surveys where galaxy positions and photometric redshifts are available. It searches for projected galaxy overdensities in photometric redshift space without any assumption on the underlying galaxy population (e.g. presence of a red sequence). \texttt{WaZP}, in a nutshell, slices the galaxy catalog in photometric redshift space and then generates smooth wavelet-based density maps for each slice with using positions (i.e. RA and Dec) of galaxies. Thus, overdensity peaks are extracted and then combined to form a distinct list of cluster candidates and associated galaxy members \citep{Aguena21}.

Following the detection of an overdensity, the radius and the richness ($\lambda$) of the clusters computed jointly. The radius of the clusters is the radius where the density is 200 times of the local galaxy background (i.e. $R_{200}$) and the richness is defined as the sum of membership probabilities of galaxies within the $R_{200}$.

Probabilities for membership for a given cluster is computed based on the distances of galaxies to the cluster center, their photometric redshifts and magnitudes \citep{Castignani16}. Details of the \texttt{WaZP} cluster finder can be found in \cite{Aguena21} which is also used in the Euclid Cluster Finder Challenge \cite{EuclidCol19}.

Cluster candidates detected by \texttt{WaZP} have a signal-to-noise (SNR) ratio computed by using the density peak relative to the local galaxy background. We have used cluster detections with a SNR $>$3 which yields 3337 detections in the whole W1 region.
\begin{figure}[t]
\begin{center}
	\includegraphics[width=\columnwidth]{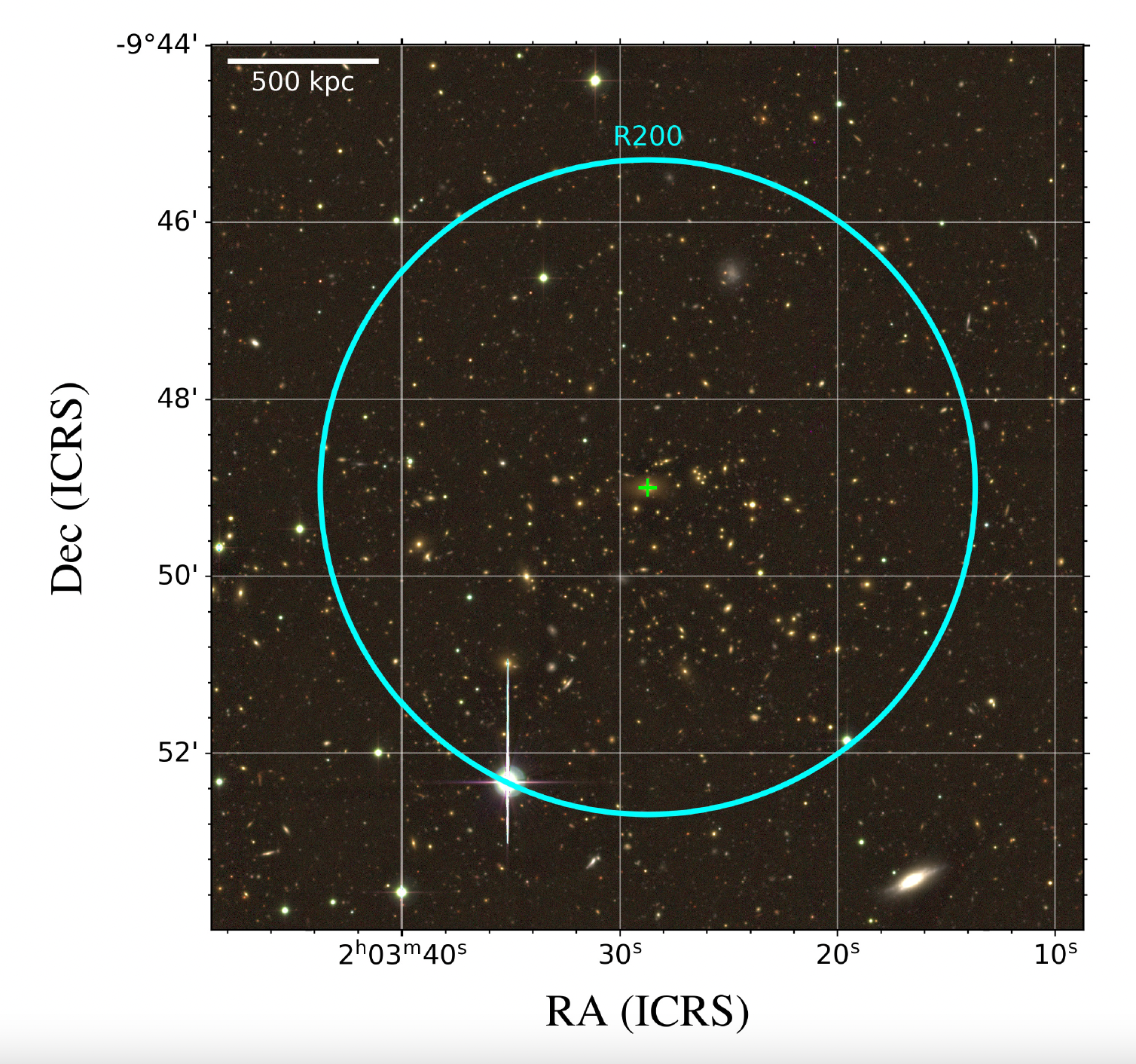}
	\caption{An example cluster field. The photometric redshift of the BCG is z=0.329. Cyan colored circle represents the $R_{200}$ radius from the cluster center. The cluster center is denoted with a cross sign.}
    \label{fig:bcgiden}
\end{center}
\end{figure}

We identified BCGs of those galaxy clusters as the brightest elliptical galaxies when the following criteria have been satisfied: i) being 0.5 Mpc around the cluster centre, ii) having a cluster consistent redshift with  $\Delta z = 0.03*(1+z_{cl})$, iii) having a $(r-i)$ color consistent (within $\pm 0.3$) with model elliptical galaxy colors for the corresponding redshift.

In order to examine cluster detections we produce true color images of the cluster cores with STIFF \citep{Bertin12} using g, r, and i-band CFHTLS images. As our cluster sample to identify BCGs is relatively modest in size, we performed a visual inspection to eliminate any false detection. We keep cluster candidates, hence BCGs, in our sample when there is a clear overdensity of galaxies with consistent colors. This leaves 3283 BCGs in W1 region.

\section{Structural Analysis of BCGs}

We use \texttt{GALFIT} for structural analysis of our sample galaxies. \texttt{GALFIT} is a 2D image decomposition tool that is used extensively for morphological studies of galaxies \citep{Peng2002}. It tries to model the surface brightness profiles of galaxies by fitting an analytical function such as de Vaucouleurs, exponential, nuker, gaussian. Amongst all these functions S\'ersic is widely used for structural analysis of galaxies. The S\'ersic function introduced by \cite{Sersic63} and \cite{Sersic1968} is considered as the generalized function where, for instance, n=1 and n=4 corresponds to the exponential and de Vaucouleurs models, respectively. Therefore, modelling the light distribution is more efficient with a single S\'ersic function.

The S\'ersic profile is given as

\begin{equation}
    I(R) = I_{e} \  exp \left(-b(n)\left[\left(\frac{R}{R_e}\right)^{1/n}-1\right]\right)
\end{equation}

\noindent where ($R_e$) is the effective radius (i.e., half-light radius), $I_{e}$ is the surface brightness at ($R_e$), $n$ is the S\'ersic index, and $b(n)$ is a dimensionless parameter that is coupled to $n$ such that half of the total flux (i.e. $I_{e}$) is always within ($R_e$).

In this study, we model $r$-band images of each BCG with a single S\'ersic profile in addition to a background (sky) model. We describe the main steps of the analysis that we performed in the following sections.

\subsection{Preparing Images}

CFHTLS W1 field consists of 72 individual MegaCam pointings (i.e. regions). For each BCG in our sample, we first determine the corresponding region by using the position of the galaxy and the corner coordinates of each region. Then, a cutout image was produced from the $r$-band image using our own \emph{CFITSIO} routines.

For the sake of the analysis with \texttt{GALFIT}, we produced cutout images of $300\times300$ pixels wide with BCGs are centered. This image size corresponds to $\sim 56$ arcseconds when MegaCam pixel scale of 0.186$^{\prime\prime}$/pixel is taken into account. At the redshift of $z\sim0.1$ this angular scale corresponds to a physical scale of $\sim100$ kpc with the standard cosmology. This physical size is more than three times larger than typical BCG sizes (e.g. 10-30 kpc) and leaves enough area for \texttt{GALFIT} to determine the background level around the galaxy.

It is important to keep in mind that \texttt{GALFIT} can fit the sky as a free parameter. Thus, a reasonable sky area is needed around the object of interest.

\subsection{Masking}

For modelling the light distribution of galaxies without the contribution of nearby sources one needs to mask other sources than the object of interest. Thus, we created a mask file to be used as an input for the GALFIT. 

Mask images for image defects, stars, spike and ghost like structures have been already created by TERAPIX for the whole W1. For each region amongst the 72 individual MegaCam pointings there is a mask file. As we determine the corresponding region for our target of interest, we produce cutout mask files for this masking with the same field-of-view of our $r$-band cutouts. Since this masking is done by using polygon shaped regions, we called this mask file as the polygon masks. However, this mask file is not enough for a reliable GALFIT solution. We also masked bright galaxies around our object of interest. This is especially relevant as we study BCGs which are residing at the cores of galaxy clusters, the most densest place in the Universe. We masked all objects other than our target galaxy (i.e. BCG) if they are brighter than $i < 20$. To determine regions to be masked we make use of the Kron radius, semi-major and semi-minor axis and the position angle of the sources. These parameters are taken from the TERAPIX object catalogs. Mask files created in this way called object masks.
\begin{figure}[t]
\begin{center}
	\includegraphics[width=\columnwidth]{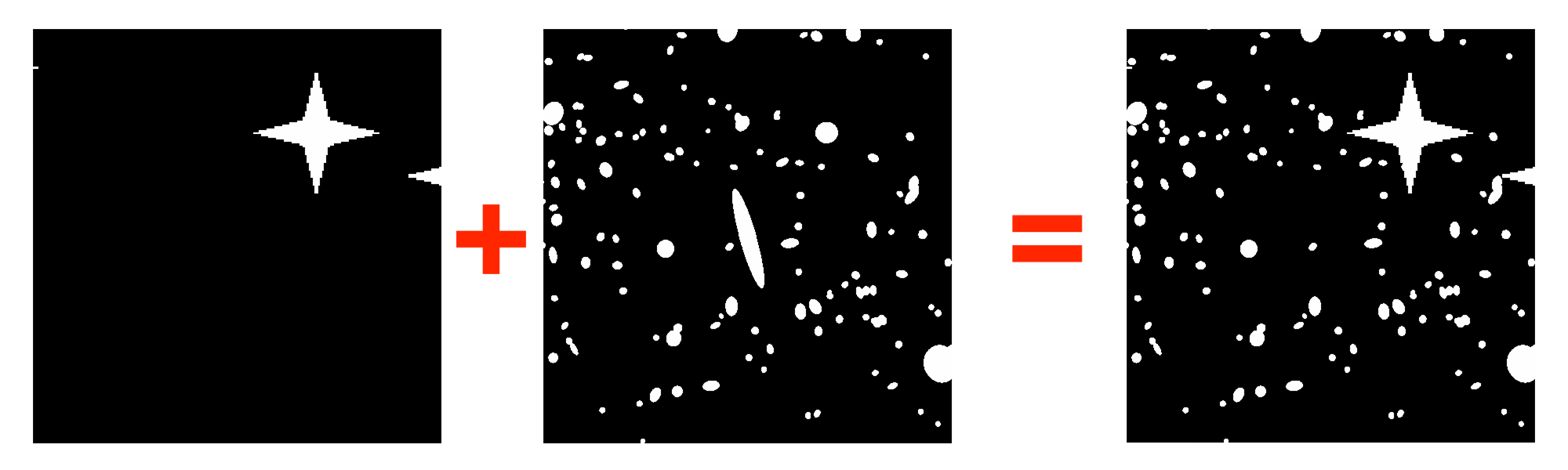}
	\caption{Masking strategy used in this study. Polygon mask file (\emph{left}) consisting of stars and similar structures provided by TERAPIX is used together with the object masks (\emph{center}) we produce to create the final mask file (\emph{right}) used by the GALFIT.}
    \label{fig:masking}
\end{center}
\end{figure}
\begin{figure}[t]
\begin{center}
	\includegraphics[width=\columnwidth]{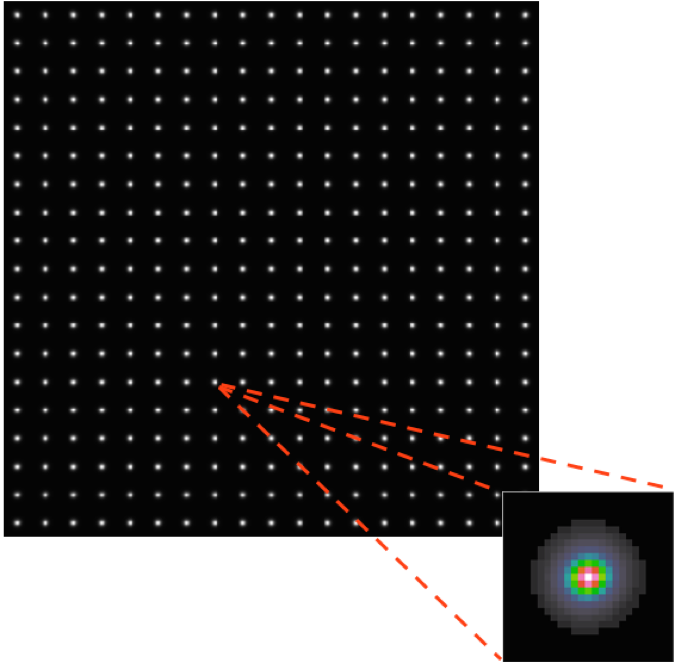}
	\caption{Point-spread function model obtained with PSFex for a single MegaCam pointing. Inset is the enlarged image of the PSF of a sub-field. PSFs are produced for the each 72 region of W1 in the similar way.}
    \label{fig:subpsf}
\end{center}
\end{figure}

Afterwards, we merge these two mask files into a final mask file to be used by GALFIT. This approach of masking and creating the final mask files is shown in Fig. \ref{fig:masking} and has been done with the similar manner given in \cite{Yelkenci2015}.
 
\subsection{Obtaining the Point-Spread Function Model}

In order to model galaxy light distributions, GALFIT requires a point-spread function (PSF) model as an input file. This is especially becomes crucial for objects residing in crowded area.

We make use of \texttt{PSFex}\footnote{https://github.com/astromatic/psfex} developed by E. Bertin in line with SExtractor. For each W1 region, we created LDAC files with SExtractor and feed them into PSFex. We let PSFex choose point-like objects automatically with imposing the maximum ellipticity of $\epsilon = 0.3$ for an object. Point-like sources are selected for the construction of the PSF if they have a signal-to-noise ratio of $S/N \geq 20$.

As PSF can vary throughout the image and each MegaCam pointing is quite large (i.e. 1 deg$^2$), we divided each W1 region (19354 $\times$ 19354 pixels) into 19 $\times$ 19 sub-fields with roughly 100 $\times$ 100 pixels each. Thus, we obtain 361 PSFs for a given MegaCam pointing. This process is repeated for the whole 72 regions in W1. 

Depending on the target's coordinates we determine the representative PSF of that sub-field to be fed into GALFIT. An example of the PSF image for a MegaCam pointing is shown in Fig. \ref{fig:subpsf} where we demonstrate choosing the relevant sub-field for a target galaxy.

\subsection{Running GALFIT}

GALFIT requires a number of input information and initial parameters for fitting the light distribution. This implies a coordinated input for running the program, especially for the case of large number of objects. Thus, we make use of GALFIT with a wrapper Fortran program and some post-processing scripts. This Fortran program enables to arrange input parameters, prepare necessary configuration and auxiliary files for the run of the whole BCG sample. Final mask and PSF files explained in the previous sections are fed into the process at this step.

The initial values of effective radius ($R_e$), axis ratio ($b/a$), position angle, and magnitude are taken from the CFHTLS object catalogs provided by TERAPIX.

\begin{figure*}[t]
\begin{center}
	\includegraphics[width=\textwidth]{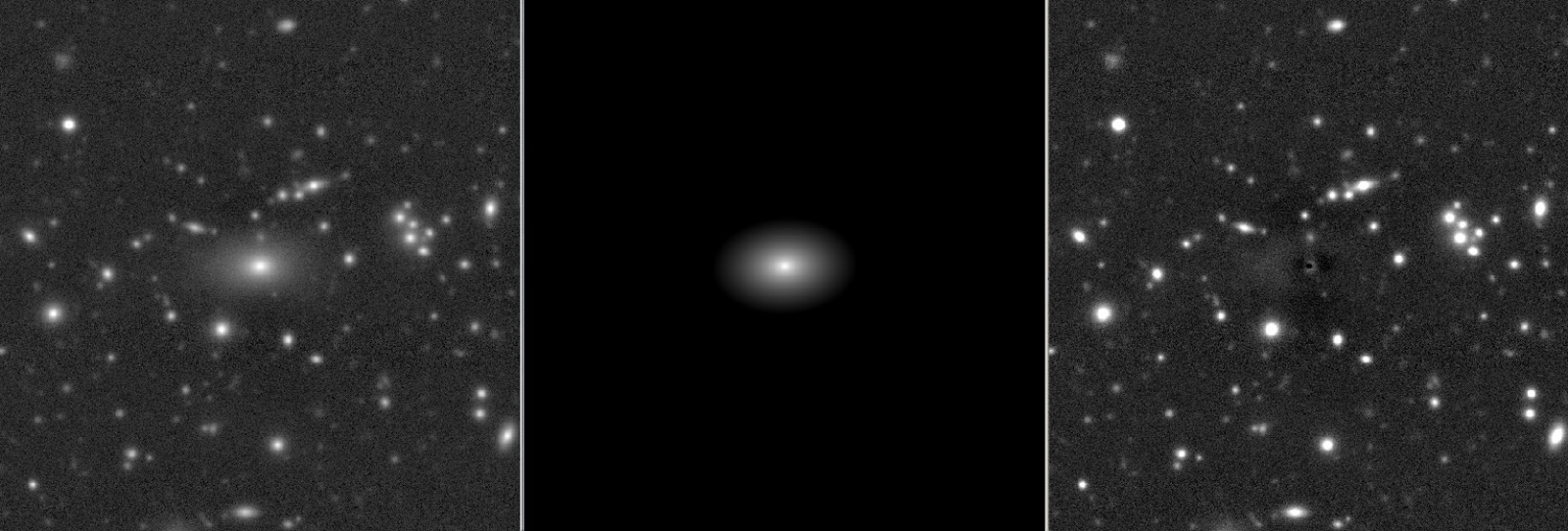}
	\caption{An example GALFIT output for the BCG ID\#7938 (zphot=0.329). Cutout (left), model (middle), and residual (right) images are shown.}
    \label{fig:galfitout}
\end{center}
\end{figure*}

The resulting output of the GALFIT run is a multi-extension FITS file which includes the input image (i.e. cutout), produced model image, and the residual image of the $(image-model)$ subtraction. In Fig. \ref{fig:galfitout} we present an example of these output images for the BCG ID\#7938.

Each individual frames shown in Fig. \ref{fig:galfitout} contains the output parameters of the fitting process as header information. Once GALFIT run is completed for an object, our wrapper program extracts these values, arranges and tabulates in an output file which is the basis of our results.

GALFIT is run in an iterative process and maximum of 100 iterations are allowed. If the fitting process is converged to a solution before the last iteration then we have the output values of fitting parameters. Otherwise, empty values are returned and our wrapper program deals with those situations to mark such cases in the output file.

Modeling the BCG light distribution requires greater care because the central regions of galaxy clusters exhibit high galaxy densities. Thus, we implemented a two-step run for GALFIT to model the sky background better. In the first run, we keep Sersic and sky components free. Not only the sky value is kept free but also the possible sky gradients along both axes of the image (i.e. dsky/dx and dsky/dy). When the first run is converged to a solution, we take the sky background and gradients and run GALFIT for a second time with those parameters fixed. The Sersic parameters of the second run are taken as the final results.
 
\subsection{Goodness of Fit}

Assessment of the quality of the model image produced by GALFIT is measured by means of $\chi^{2}$ via Levenberg-Marquardt algorithm. Although initial values of the free parameters is given as meaningful as possible, a large range of values is inspected during the fitting process and Levenberg-Marquardt algorithm is the optimal way for searching the best values. GALFIT continues fitting with an iterative approach until the $\chi^{2}$ does not change significantly.

The goodness of the light profile fitting is then given as the reduced $\chi^{2}_{\nu}$ \citep{Peng2010}:

\begin{equation}
  \chi^{2}_{\nu} = \frac {1}{N_{dof}} \sum_{x=1}^{nx} \sum_{y=1}^{ny} \frac{{(f_{data} (x, y)-f_{model}(x, y))^{2}}}{\sigma{(x, y)}^{2}}
\end{equation}

\noindent where ($N_{dof}$) is the number of degrees of freedom, $f_{data} (x,y)$ and $f_{model} (x,y)$ are the input and the model images, respectively. {$\sigma (x, y)$} is the 2D data for the uncertainty on the flux either created internally or given as an input. In our case, we let GALFIT to produce the relevant $\sigma$-image by using the \emph{GAIN} and the \emph{READ NOISE} parameters.

$f_{model}(x, y)$ is the sum of analytical functions that are the product of a number of free parameters. In this study, we used only S\'ersic function to model the data and we let centroid position, integrated (total) magnitude, effective radius, S\'ersic index, axis ratio, and position angle as free parameters.

\section{Results and Discussion}

Our pipeline followed the procedures explained in the previous section and could converge to a solution for 2721 BCGs. Before continuing with analysis we wanted to remove nonphysical or unreliable solutions from our results. Thus, we applied the following criteria for the GALFIT outputs; $\chi^{2}<5$, $n<8$, and $R_{e}<50 \ kpc$. Larger S\'ersic indices are not reliable hence we omit the ones larger than eight. For the effective radius of BCGs, larger values than 50 kpc are also not easily motivated physically, therefore we also omit those cases. 

\begin{figure}[t]
\begin{center}
	\includegraphics[width=9cm]{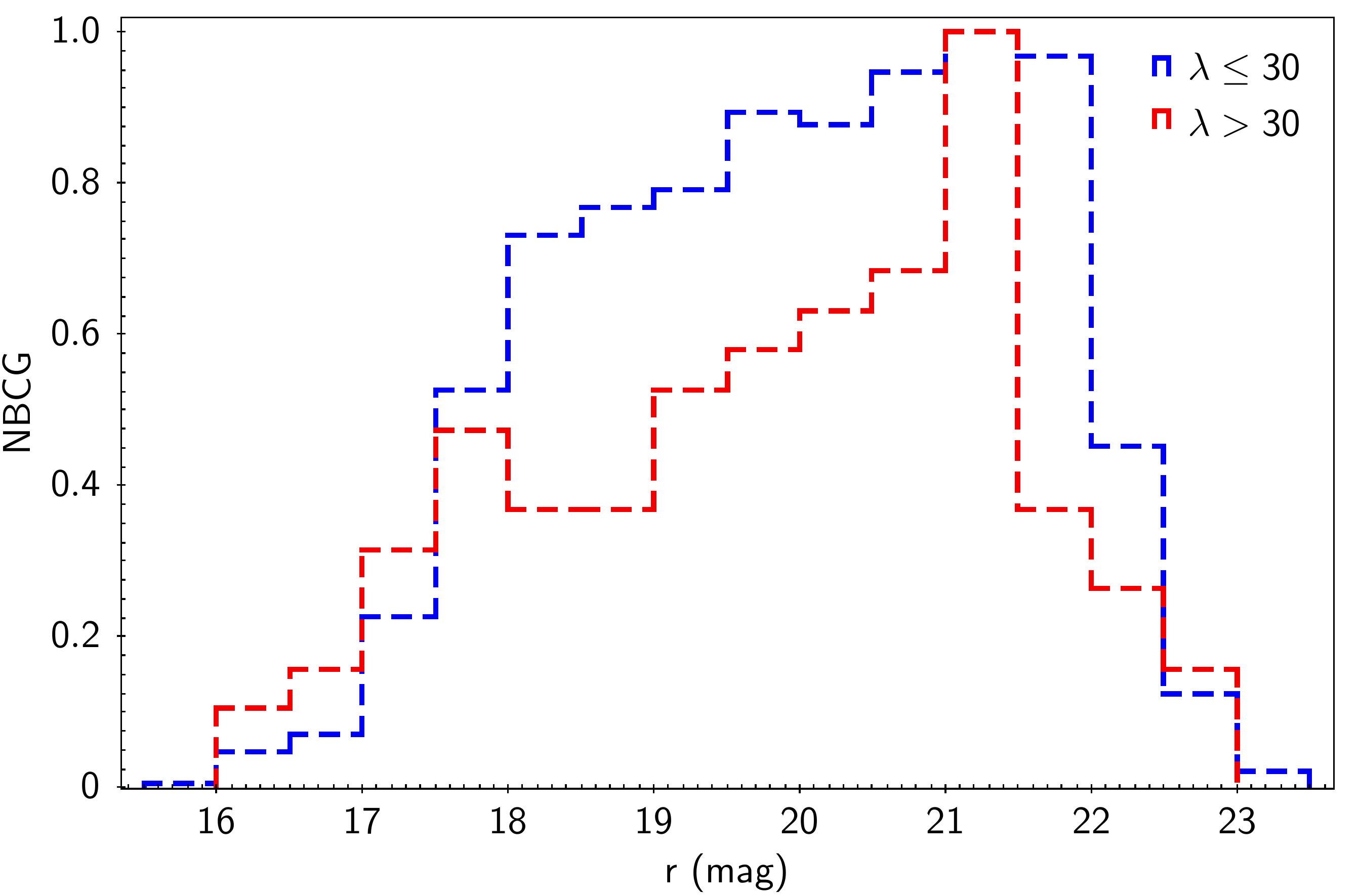}
	\includegraphics[width=9cm]{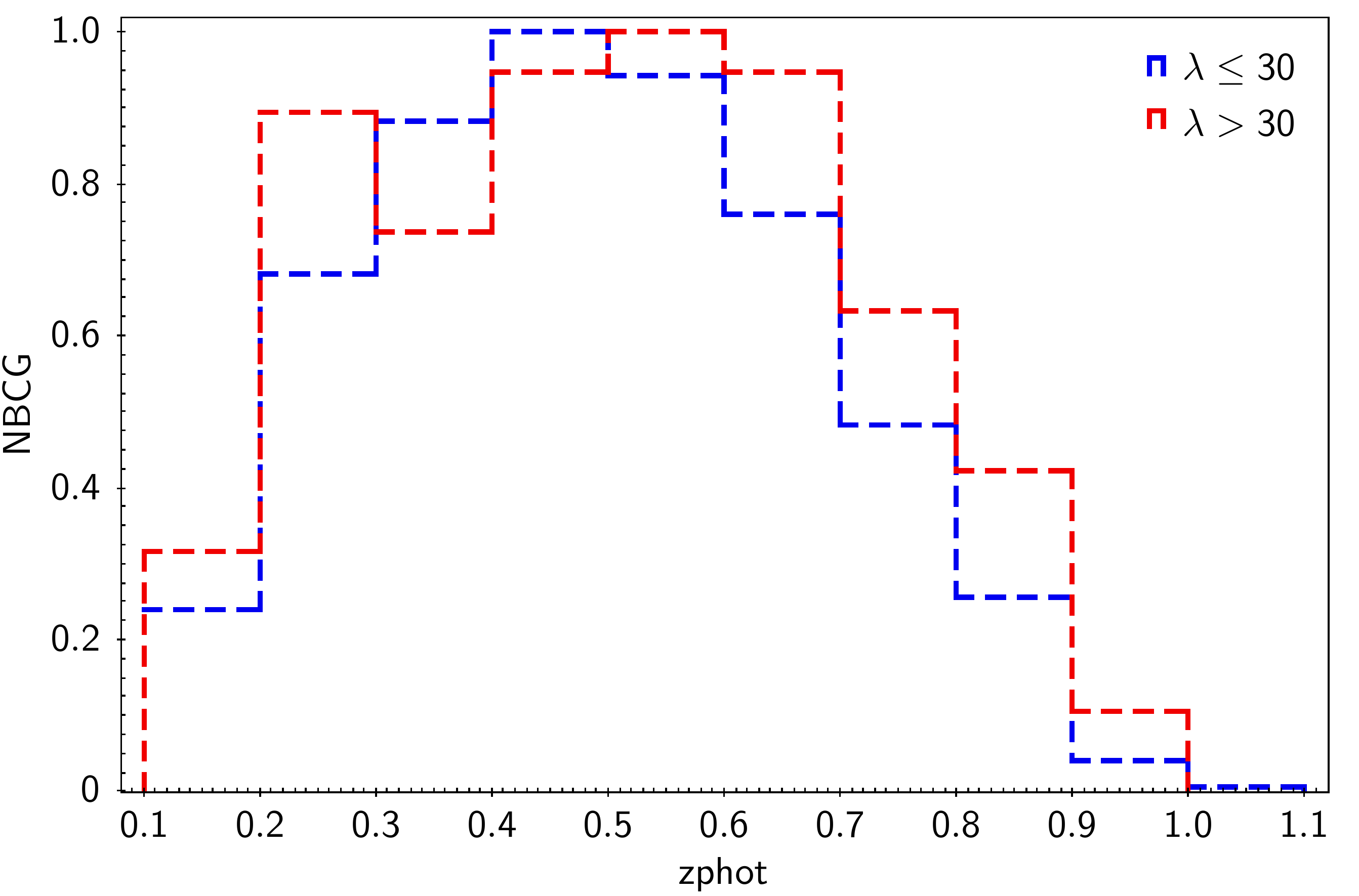}
	\caption{Normalized distributions of $r$-band apparent magnitude (top) and photometric redshift (bottom) of BCGs. In each histogram the whole sample divided into two subsamples as for the lower richness ($\lambda \leq 30$) and higher richness ($\lambda > 30$) clusters.}
    \label{fig:histo_mag_zphot}
\end{center}
\end{figure}

One final cut has been applied to the redshift range. As we have a few BCGs at lower ($z<0.1$) and higher ($z>1.0$) redshifts, we did not include them for the following analysis of the scaling relations. Thus, 1685 BCGs are left after this elimination within the redshift range of $0.1 < z \leq 1.0$. The distributions of $r$-band magnitudes and the photometric redshifts for the final sample are given in Fig. \ref{fig:histo_mag_zphot}.

In the following sections, we analyzed the structural parameters according to the host cluster richness ($\lambda>30$ and $\lambda \leq 30$) and the redshift of the BCG ($0.1<z\leq0.4$, $0.4<z\leq0.7$, and $0.7<z\leq1.0$). Statistics of the structural parameters (i.e. $R_{e}$ and $n$) are given in Table \ref{tab:stats_logRe_n} for different richness and redshift bins.

We provide relations for parameter pairs of the "photometric plane" (PP) of early-type galaxies in which velocity dispersion is replaced with S\'ersic index. The PP links $log(R_{e})$, $\mu_{e}$, and $log(n)$ to constitute a three-dimensional relation for early-type galaxies \citep{Barbera2004}.

\subsection{Distribution of structural parameters}

\begin{figure*}[t]
\begin{center}
	\includegraphics[width=\columnwidth]{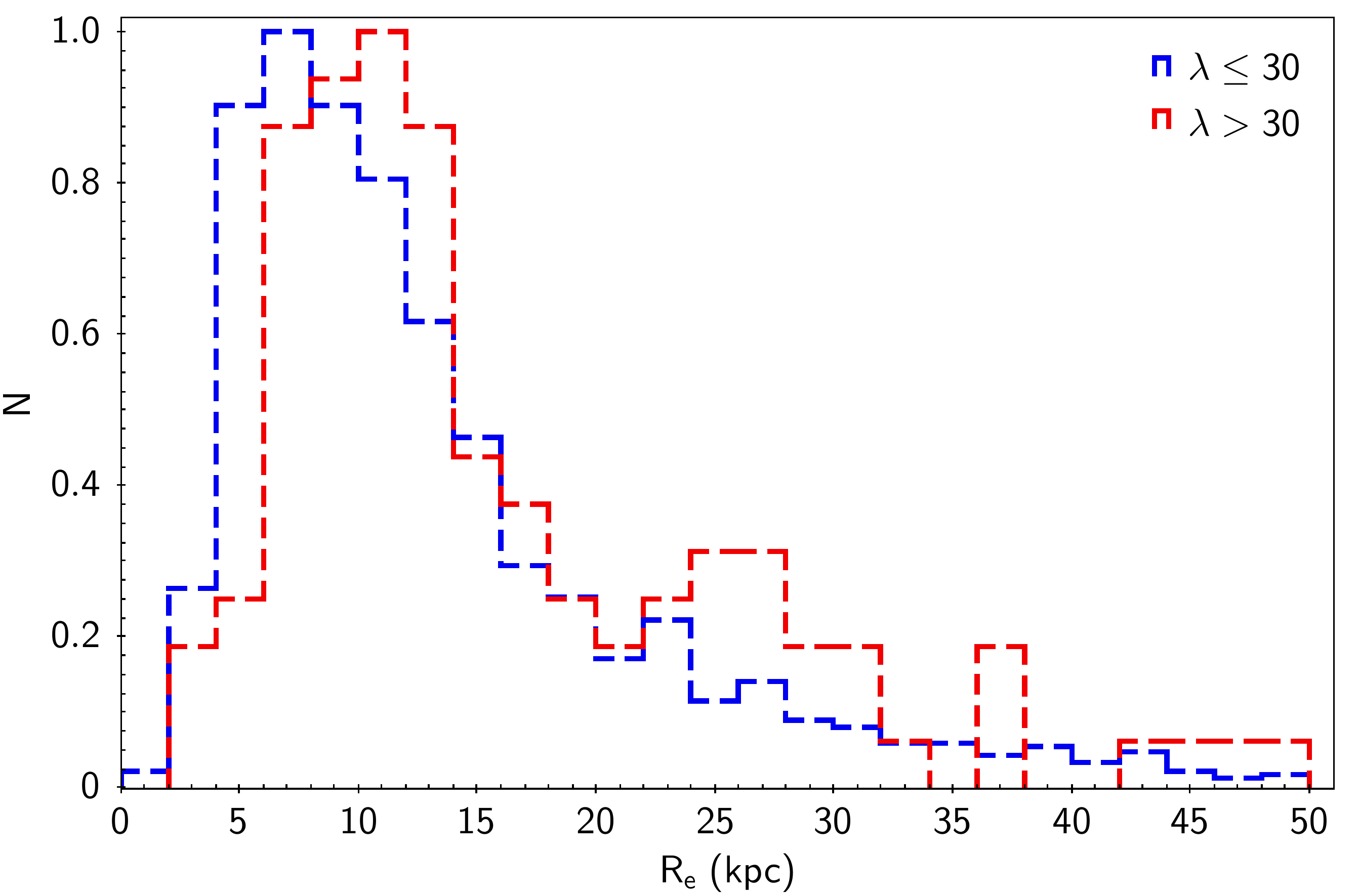}
	\includegraphics[width=\columnwidth]{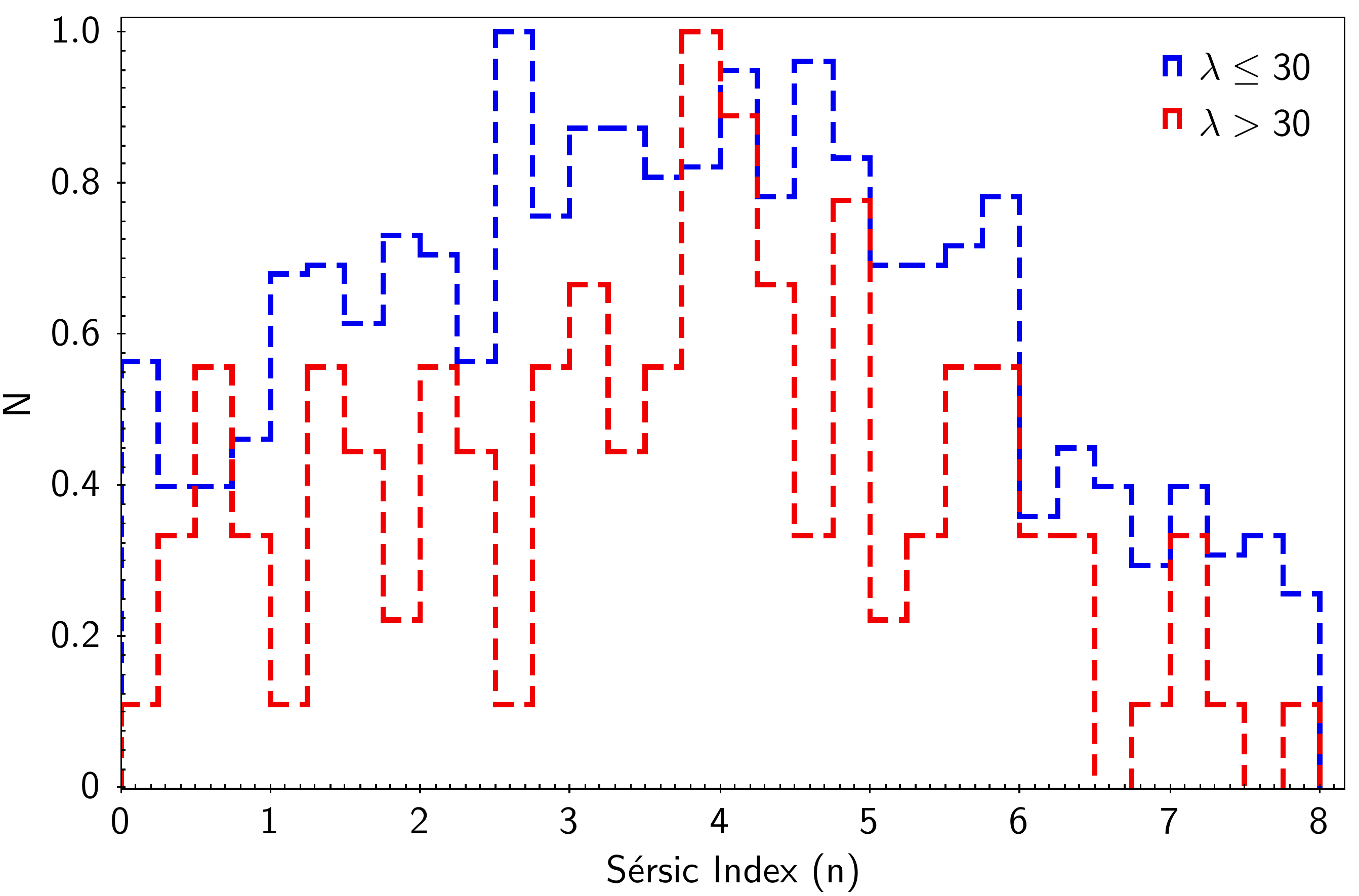}
	\caption{Normalized distribution of S\'ersic index and the effective radius for the whole redshift range but for different richness cuts.}
    \label{fig:histo_structural}
\end{center}
\end{figure*}

We compared the structural parameters of BCGs according to the host cluster richness in Fig. \ref{fig:histo_structural}. The effective radius distribution of rich clusters is skewed towards higher radii which implies that BCGs residing in richer clusters tend to have larger effective radius. This behaviour exists at all redshift bins and can be seen from the median values of effective radius given in Table \ref{tab:stats_logRe_n}. However, this difference between relatively poor and rich clusters become more evident in the lowest redshift bin which can be an implication of the more merging events in denser environments. On the other hand, S\'ersic index of our BCGs both in poor and rich clusters show a similar distribution.

We performed statistical tests to check whether BCGs from poor and rich clusters are drawn from the same sample. For the Sersic index, Kolmogorov-Smirnov (KS) and ANOVA tests reveal the p-values as 0.8 and 0.7, respectively. These results suggest that it is difficult to distinguish these two samples by their S\'ersic index.

On the other hand, effective radii of BCGs in poor and rich clusters are statistically different based on their p-values of $6.6\times10^{-4}$ and $6.7\times10^{-4}$ for KS and ANOVA tests, respectively. The latter result is consistent with the results of \cite{Ascaso2011} which shows significant correlations between the X-ray luminosity of the host cluster and the absolute magnitude of the BCG. The X-ray luminosity, hence the mass of the cluster \citep{Vikhlinin2006} is indicated with the cluster richness in our study. Since BCGs follow a size-luminosity relation as the other early-type galaxies \citep{Ulgen2022,Samir2020}, absolute magnitude of the BCGs can be correlated with the effective radius determined in this study.
\begin{figure*}[t]
\begin{center}
	\includegraphics[width=\columnwidth]{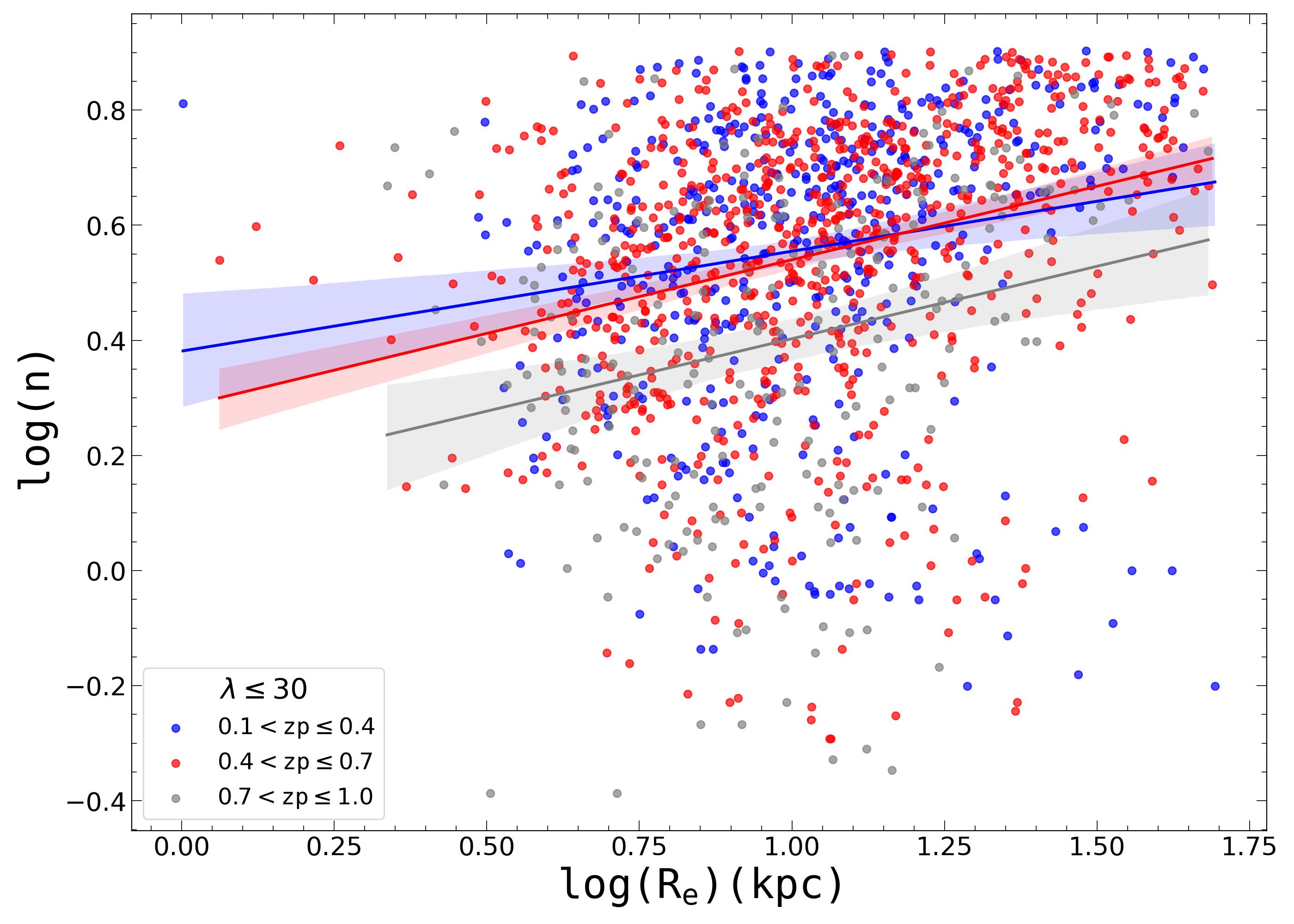}
	\includegraphics[width=\columnwidth]{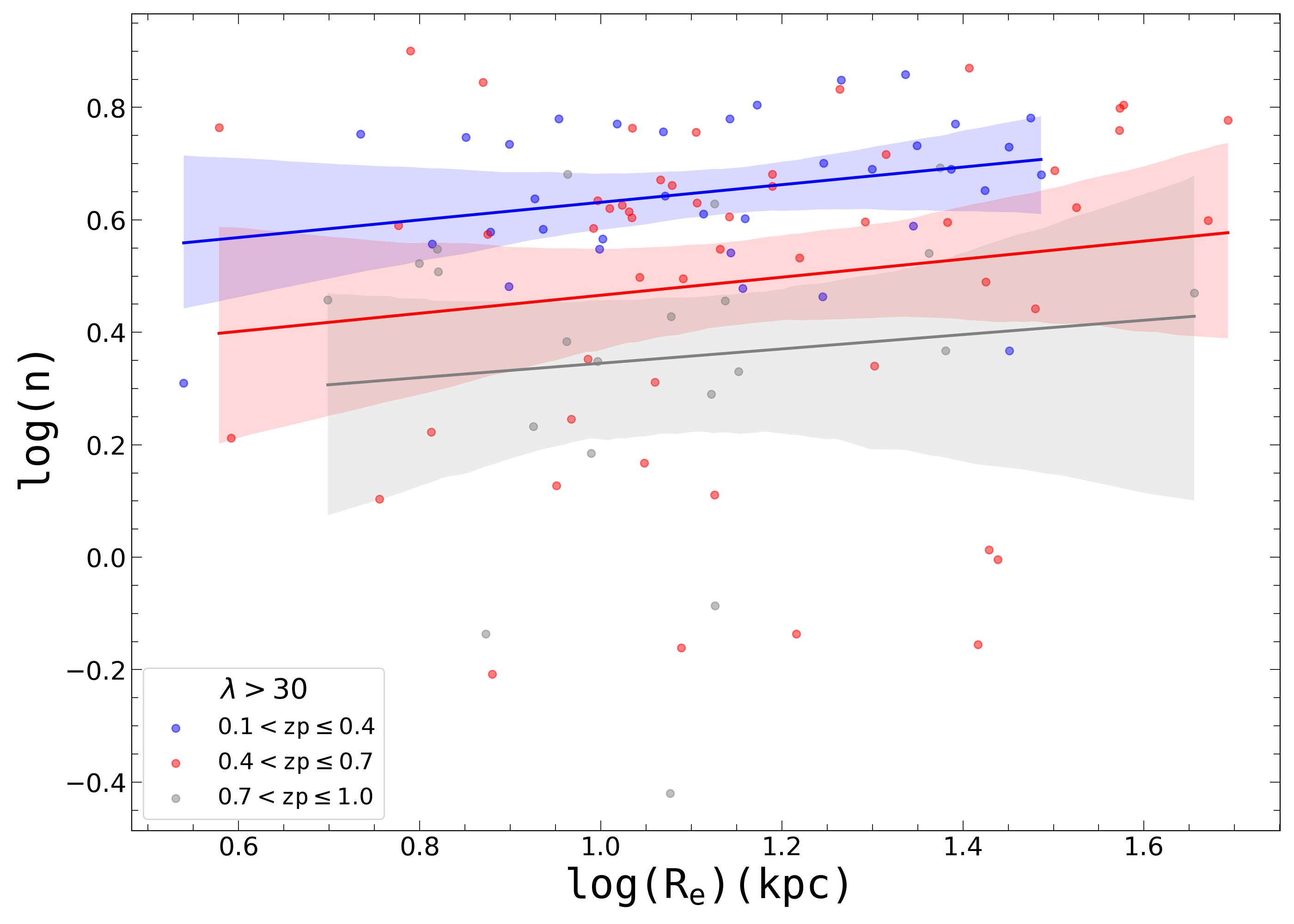}
	\caption{Relation between logRe-logn for poor (left) and rich (right) clusters. Blue, red, and gray points and associated linear fits represent different redshift bins. For each fit $2\sigma$ confidence intervals are also displayed.}
    \label{fig:logRe_logn}
\end{center}
\end{figure*}

In Fig. \ref{fig:logRe_logn} we show the relation between effective radius ($R_{e}$) and S\'ersic index (n). With a large scatter, the trend is that larger BCGs have larger S\'ersic indices which is also shown by \cite{Chu2022}. Equation \ref{eq:logRe_logn} provides the best-fit relation that is obtained from the whole sample.
\begin{equation}
log (n) = 0.231 \ \times \ log (R_e) + 0.291   
\label{eq:logRe_logn}
\end{equation}

Table \ref{tab:logRe_logn} lists results for individual redshift bins and for different host cluster richnesses. The slope of the whole sample, irrespective of redshift or richness, is consistent with those of \cite{Ascaso2011} where they provided the best-fit relations for BCG samples at $z \sim 0$ and $z \sim 0.5$.

$log R_{e} - log(n)$ relation shows a common behavior for all redshift bins with slopes between 0.2-0.3. However, the slopes for the BCGs in rich clusters are less steep.
\begin{table*}[t]
\centering
\caption{Statistical properties (number of objects, mean, standard deviation ($\sigma$) and median) of effective radius ($R_{e}$) (in kpc) and S\'ersic index of the BCGs for different richness and redshift bins.}
\small
\begin{tabular} {@{}c|cccc|cccc|cccc@{}}
\hline
\noalign{\vskip 1mm}
\multicolumn{13}{c}{Effective Radius ($\text{R}_e$) (kpc)} \\
\noalign{\vskip 1mm}
\hline
\noalign{\vskip 1mm}
\multicolumn{1}{c}{} & \multicolumn{4}{c}{$0.1<z\leq0.4$} & \multicolumn{4}{c}{$0.4<z\leq0.7$} & \multicolumn{4}{c}{$0.7<z\leq1.0$}\\
\noalign{\vskip 1mm}
\hline
\noalign{\vskip 1mm}
 Richness & N & Mean & $\sigma$ & Median & N & Mean & $\sigma$ & Median & N & Mean & $\sigma$ & Median\\
\noalign{\vskip 1mm}
\hline
$\lambda\leq30$ & 535 & 13.074 & 8.489 & 10.896 & 803 & 13.745 & 9.321 & 10.985 & 231 & 11.376 & 7.606 & 9.084 \\
$\lambda>30$    & 37 & 15.399 & 7.520 & 13.910 & 55 & 17.057 & 10.669 & 12.767 & 22 & 14.923 & 10.525 & 11.946\\
\hline
\noalign{\vskip 1mm}
\multicolumn{13}{c}{S\'ersic Index (n)} \\
\noalign{\vskip 1mm}
\hline
\noalign{\vskip 1mm}
\multicolumn{1}{c}{} & \multicolumn{4}{c}{$0.1<z\leq0.4$} & \multicolumn{4}{c}{$0.4<z\leq0.7$} & \multicolumn{4}{c}{$0.7<z\leq1.0$}\\
\noalign{\vskip 1mm}
\hline
\noalign{\vskip 1mm}
Richness   & N & Mean & $\sigma$ & Median & N & Mean & $\sigma$ & Median & N & Mean & $\sigma$ & Median \\
\noalign{\vskip 1mm}
\hline
$\lambda\leq30$ & 535 & 3.968 & 2.011 & 4.070 & 803 & 3.847 & 1.959 & 3.830 & 231 & 2.857 & 1.775 & 2.590 \\
$\lambda>30$ & 37 & 4.480 & 1.494 & 4.490 & 55 & 3.596 & 1.973 & 3.940 & 22 & 2.511 & 1.282 & 2.550 \\
\hline
\end{tabular}
\label{tab:stats_logRe_n}
\end{table*}
\begin{table*}[t]
\centering
\caption{Best-fit parameters for the relation between $log (R_e)-log(n)$ for different redshift and richness bins where a is the slope, and b is the intercept.}
\begin{tabular}{@{}c|cc|cc|cc|cc@{}}
\hline
\noalign{\vskip 1mm}
\multicolumn{1}{c}{} & \multicolumn{2}{c}{All redshift} & \multicolumn{2}{c}{$0.1<z\leq0.4$} & \multicolumn{2}{c}{$0.4<z\leq0.7$} & \multicolumn{2}{c}{$0.7<z\leq1.0$}\\
\noalign{\vskip 1mm}
\hline
\noalign{\vskip 1mm}
     & \emph{a} & \emph{b} & \emph{a} & \emph{b} & \emph{a} & \emph{b} & \emph{a} & \emph{b} \\
\noalign{\vskip 1mm}
\hline
 All  & 0.231 & 0.291 & 0.187 & 0.373 & 0.241 & 0.295 & 0.235 & 0.162 \\
$\lambda\leq30$ & 0.236 & 0.288 & 0.178 & 0.377 & 0.256 & 0.284 & 0.248 & 0.156 \\
$\lambda>30$ & 0.155 & 0.348 & 0.155 & 0.475 & 0.162 & 0.308 & 0.101 & 0.260 \\
\hline
\end{tabular}
\label{tab:logRe_logn}
\end{table*}
\begin{table*}[t]
\centering
\caption{Best-fit parameters for the relation between $log(R_e) \ - <\mu_e>$ (i.e. Kormendy) for different redshift and richness bins where a is the slope, and b is the intercept.} 
\begin{tabular}{@{}c|cc|cc|cc|cc@{}}
\hline
\noalign{\vskip 1mm}
\multicolumn{1}{c}{} & \multicolumn{2}{c}{All redshift} & \multicolumn{2}{c}{$0.1<z\leq0.4$} & \multicolumn{2}{c}{$0.4<z\leq0.7$} & \multicolumn{2}{c}{$0.7<z\leq1.0$}\\
\noalign{\vskip 1mm}
\hline
\noalign{\vskip 1mm}
 & \emph{a} & \emph{b} & \emph{a} & \emph{b} & \emph{a} & \emph{b} & \emph{a} & \emph{b} \\
\noalign{\vskip 1mm}
\hline
 All            & 3.941 & 18.955 & 4.005 & 17.933 & 4.062 & 19.141 & 3.863 & 20.222 \\
$\lambda\leq30$ & 3.958 & 18.940 & 4.035 & 17.907 & 4.075 & 19.135 & 3.972 & 20.122 \\
$\lambda>30$    & 3.546 & 19.381 & 3.000 & 18.951 & 4.055 & 19.089 & 3.203 & 20.822 \\
\hline
\end{tabular}
\label{tab:kormendy}
\end{table*}

\subsection{The Kormendy relation}

The Kormendy relation correlates the effective radius of an early-type galaxy with its mean surface brightness within the same radius \citep{Kormendy1977}. The relation has the form of $\langle \mu \rangle_{e} = a \times log(R_{e}) + b$ where a and b are the slope and the intercept, respectively.

Since the Kormendy relation is a projection of the Fundamental Plane (FP) of early-type galaxies \citep{Dressler1987}, it provides information on the size evolution of galaxies \citep{Longhetti2007,Tortorelli2018} and clues about their formation \citep{Kormendy2009}.

\begin{figure*}[t]
\begin{center}
	\includegraphics[width=\columnwidth]{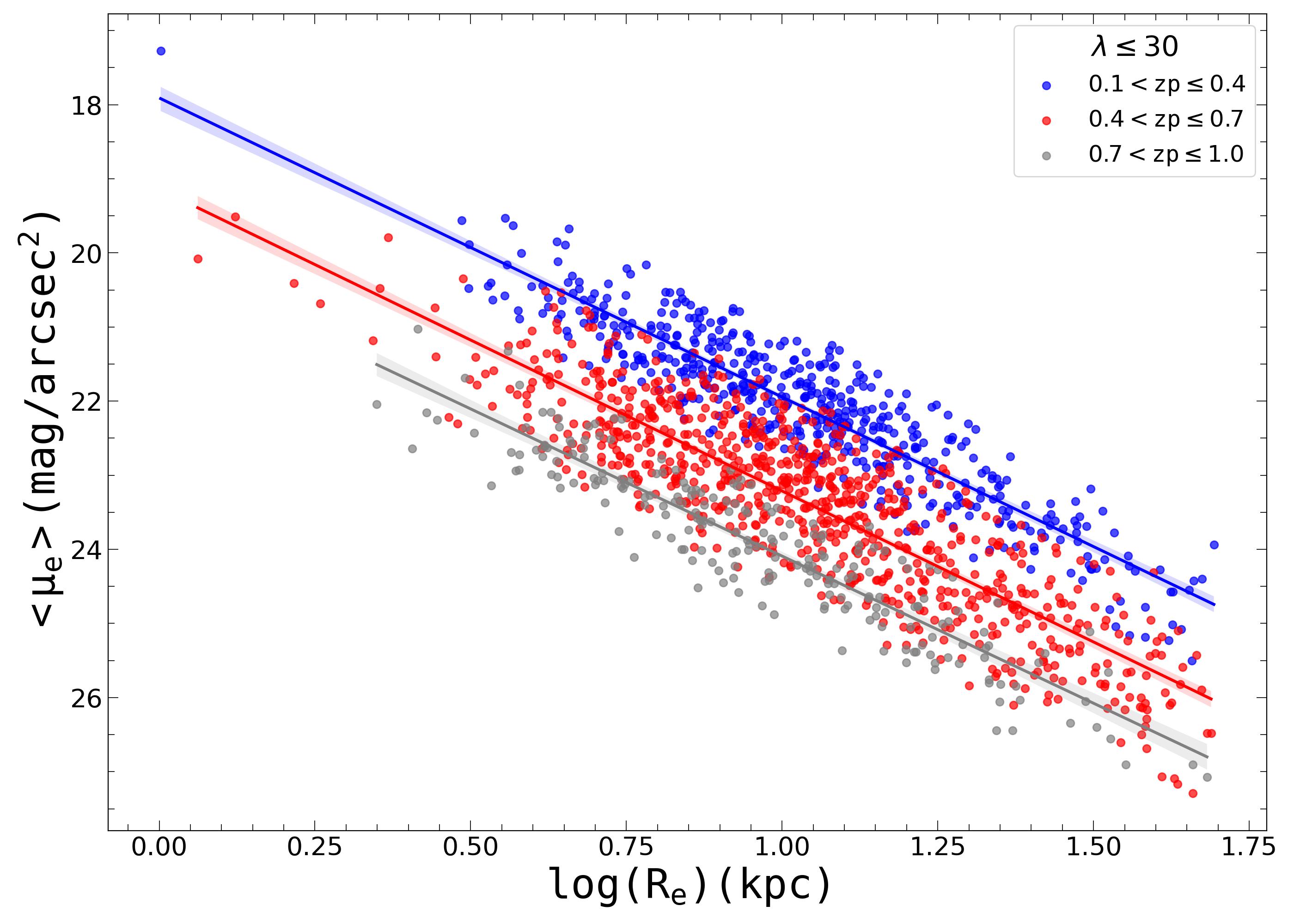} \includegraphics[width=\columnwidth]{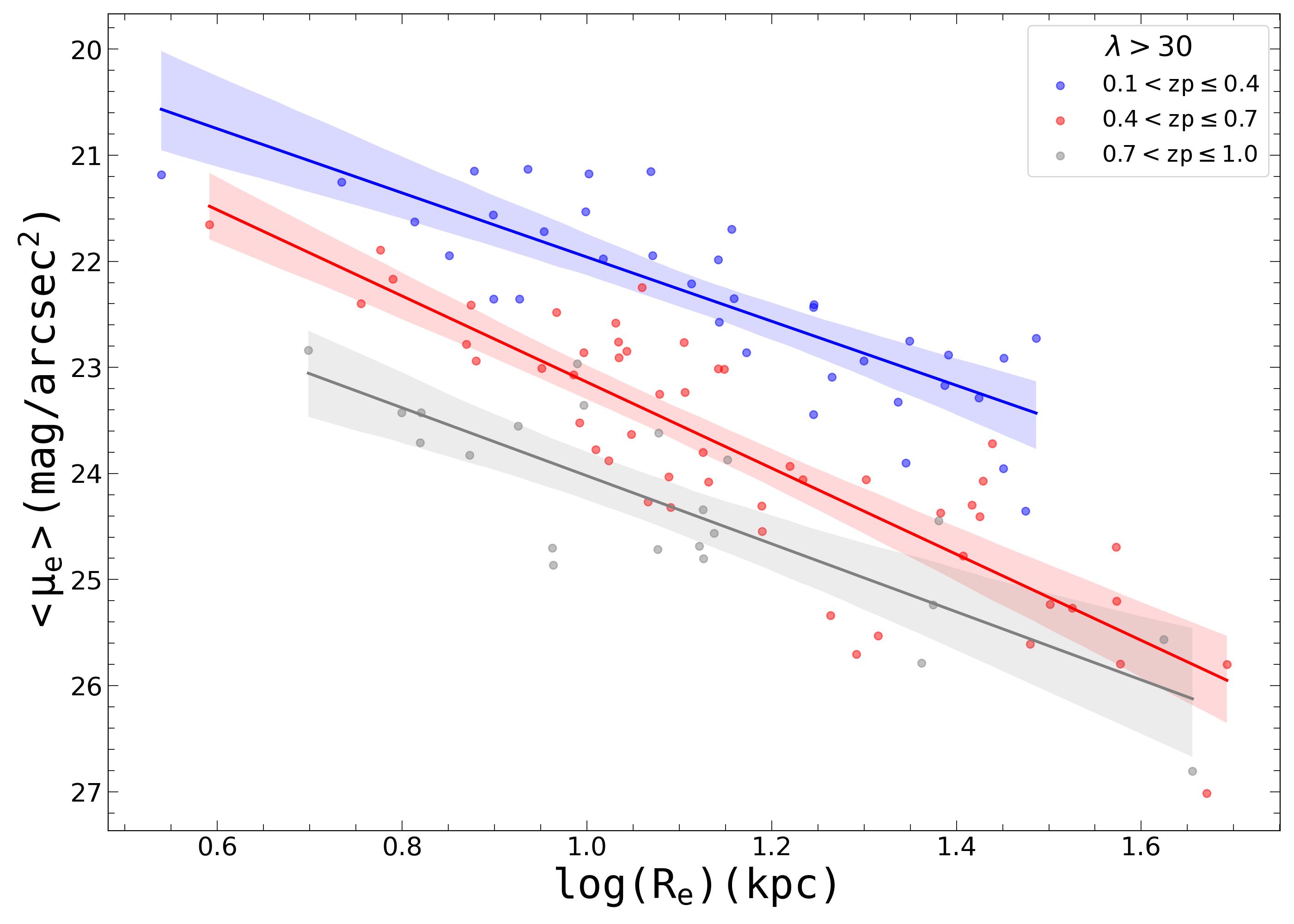} 
	\caption{Kormendy relation for poor (left) and rich (right) clusters. Points, lines, and shaded areas have the same meaning as in Fig. \ref{fig:logRe_logn}.}
    \label{fig:kormendy}
\end{center}
\end{figure*}

To obtain Kormendy relation we computed the mean surface brightness within the effective radius as given by \cite{GrahamDriver2005}:
\begin{equation}
\langle \mu \rangle_{e} = M_{tot}+2.5\ log(2\pi R_e^2) + 2.5\ log(b/a)
	\label{eq:mu_e}
\end{equation}
where $M_{tot}$ is the integrated (total) magnitude, $R_{e}$ is the effective radius, and (b/a) is the axis ratio with a and b are semimajor and semiminor axis, respectively. In Fig. \ref{fig:kormendy} we present the Kormendy relation for our BCGs both in poor and rich clusters in different redshift bins. 

We obtain a Kormendy relation irrespective of the redshift and the richness bins as given below:
\begin{equation}
\mu_e  = 3.941 \times log (R_e) + 18.955
\label{eq:kormendy}
\end{equation}

The coefficients of the Kormendy relations for different redshift and richness bins are given in Table \ref{tab:kormendy}.

\begin{table*}[t]
\centering
\caption{Best-fit parameters for the size-luminosity relation for different redshift and richness bins where a is the slope, and b is the intercept.} 
\begin{tabular}{@{}c|cc|cc|cc|cc@{}}
\hline
\noalign{\vskip 1mm}
\multicolumn{1}{c}{} & \multicolumn{2}{c}{All redshift} & \multicolumn{2}{c}{$0.1<z\leq0.4$} & \multicolumn{2}{c}{$0.4<z\leq0.7$} & \multicolumn{2}{c}{$0.7<z\leq1.0$}\\
\noalign{\vskip 1mm}
\hline
\noalign{\vskip 1mm}
 & \emph{a} & \emph{b} & \emph{a} & \emph{b} & \emph{a} & \emph{b} & \emph{a} & \emph{b} \\
\noalign{\vskip 1mm}
\hline
 All            & -0.240 & -4.481 & -0.263 & -4.961 & -0.290 & -5.629 & -0.256 & -4.990 \\
$\lambda\leq30$ & -0.242 & -4.519 & -0.265 & -5.022 & -0.290 & -5.631 & -0.259 & -5.060 \\
$\lambda>30$    & -0.196 & -3.422 & -0.236 & -4.311 & -0.274 & -5.213 & -0.261 & -5.110 \\
\hline
\end{tabular}
\label{tab:size-luminosity}
\end{table*}

\subsection{Size-luminosity relation}

\begin{figure*}[ht]
\begin{center}
	\includegraphics[width=\columnwidth]{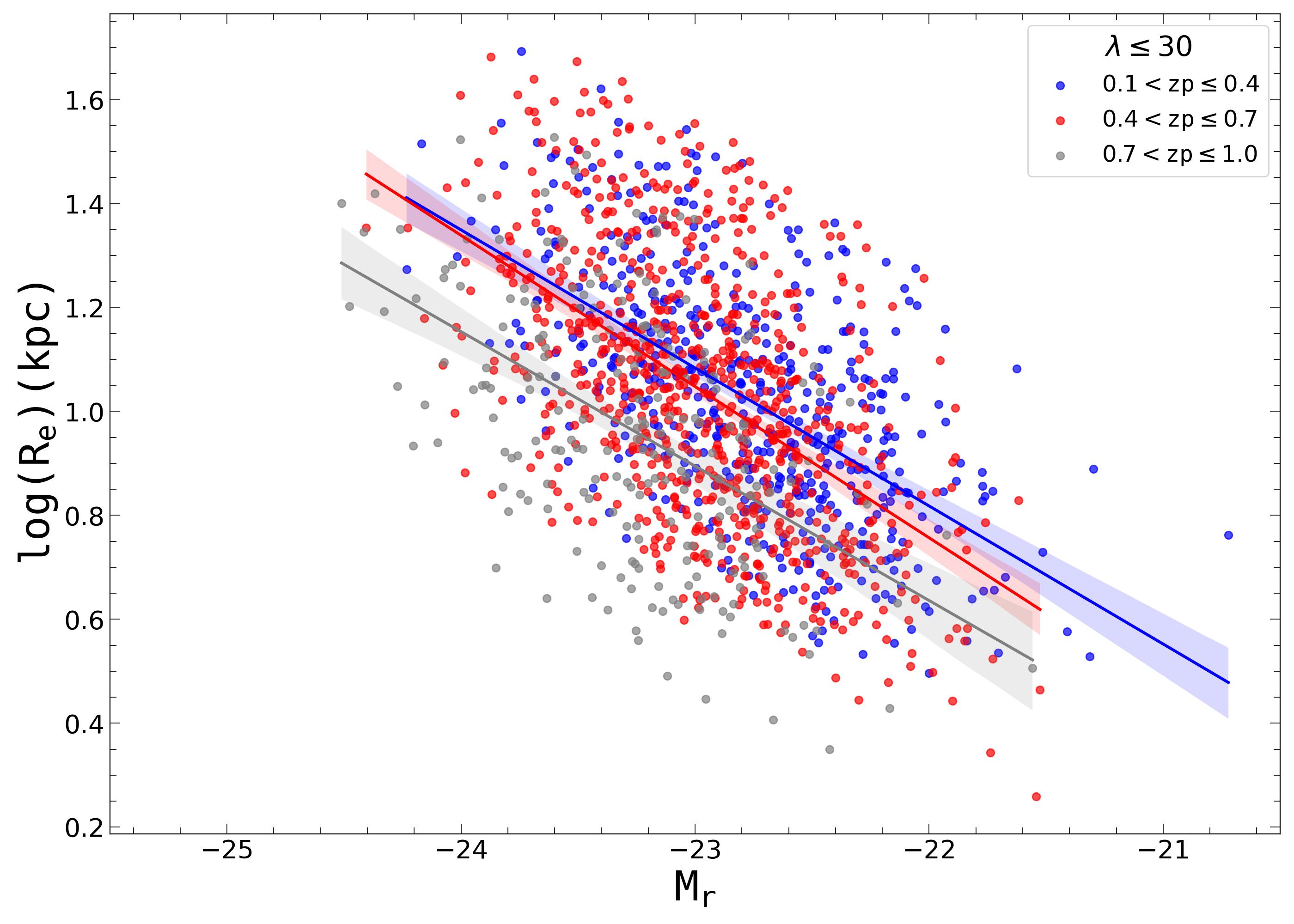}
	\includegraphics[width=\columnwidth]{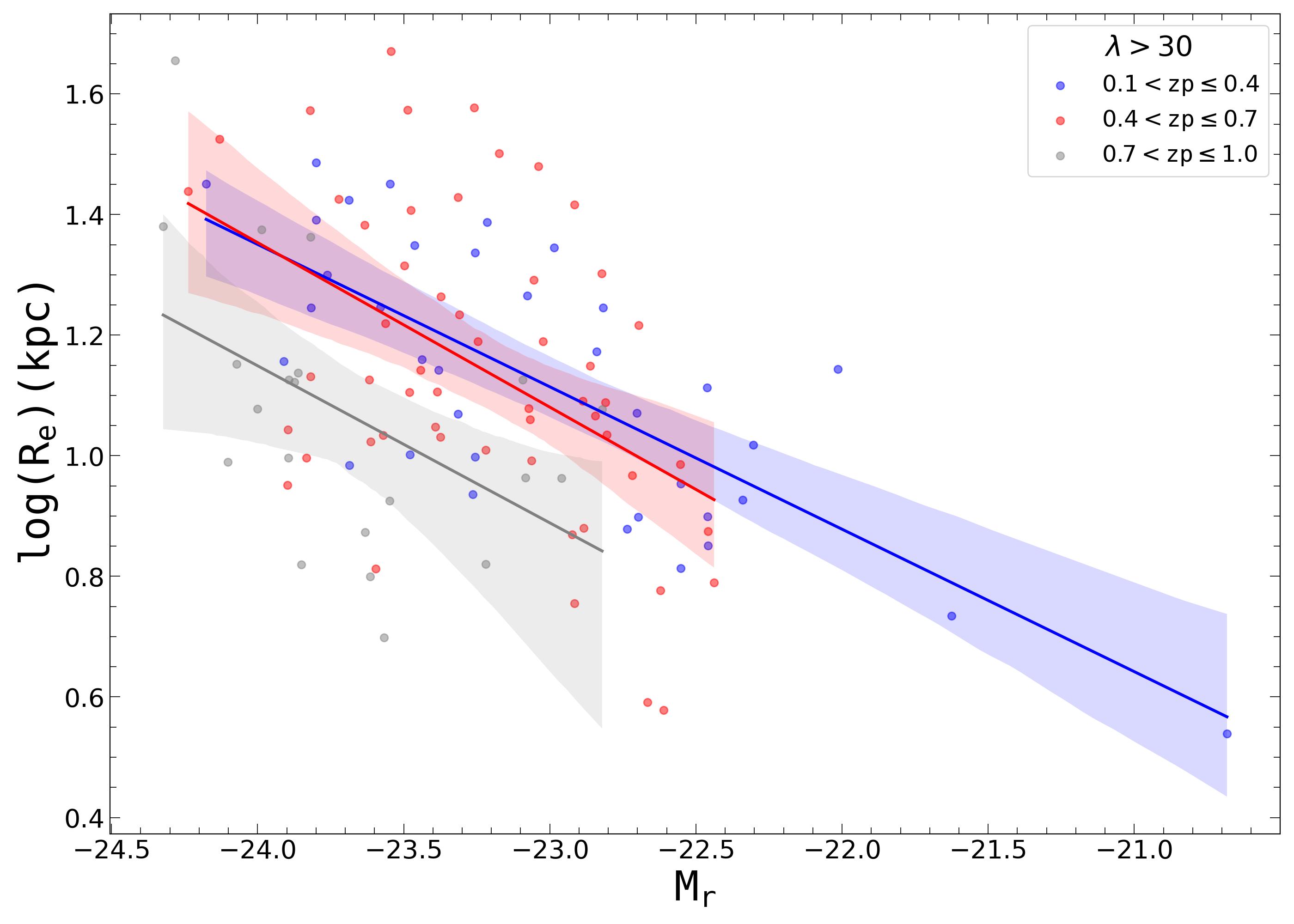}
	\caption{Size-luminosity for poor (left) and rich (right) clusters. Points, lines, and shaded areas have the same meaning as in Fig. \ref{fig:logRe_logn}.}
    \label{fig:size-luminosity}
\end{center}
\end{figure*}

We present the absolute magnitude versus effective radius relation in Figure \ref{fig:size-luminosity}. The correlation of $r$-band absolute magnitude ($\textrm M_{r}$) with effective radius seems very similar in all redshift bins. Results of the linear fits are also displayed in Fig. \ref{fig:size-luminosity}.

We obtain a size-luminosity relation for the whole sample as given below:
\begin{equation}
log  (R_e) = -0.240 \times\ M_{r} - 4.481 
\label{eq:size-luminosity}
\end{equation}

The coefficients for best-fits of the other bins are given in Table \ref{tab:size-luminosity}. The slope of the size-luminosity relation in the highest redshift bin is slightly larger for both poor and rich clusters. However, the difference in slopes of the redshift bins is not very significant.

\cite{Samir2020} also showed a linear relation between the BCG sizes and their luminosities from the sample they drawn from SDSS. While we prefer to present absolute magnitudes, we convert our magnitudes to luminosities to obtain $log\ L_r - log\ R_e$ similar to them. Since they do not present their results with a similar fashion of our study, for this comparison we did not separate BCGs of poor and rich clusters. The slope of our best-fit linear relation for the whole sample is a=0.33$\pm$0.02 whereas \cite{Samir2020} found the slope as a=0.72$\pm$0.02. The different redshift ranges of the two studies might be the reason for the slope difference.

\subsection{Evolution of structural parameters}
\begin{figure}[t]
\begin{center}
	\includegraphics[width=9cm]{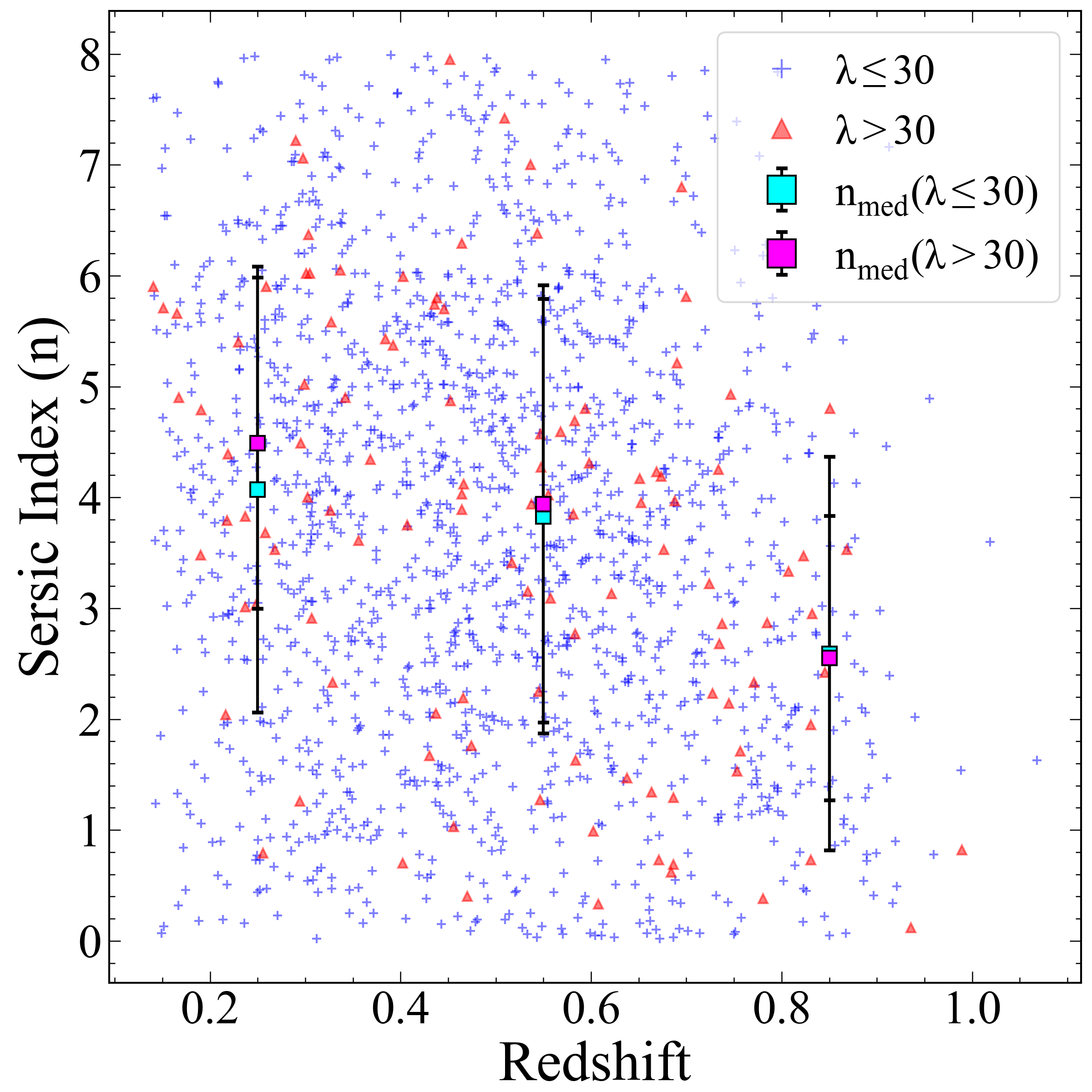}
	\includegraphics[width=9cm]{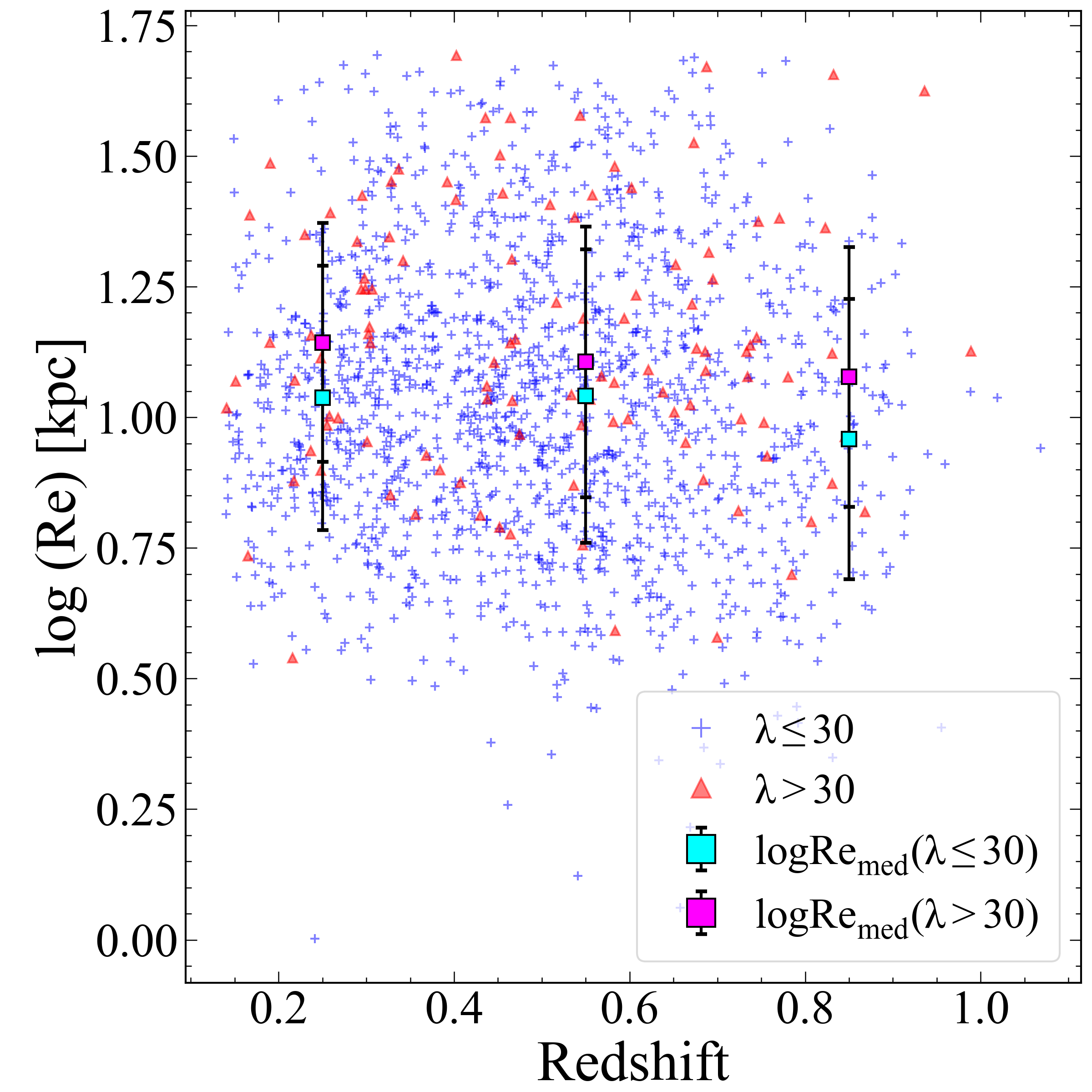}
	\caption{Evolution of the S\'ersic index (top) and the effective radius (bottom) for poor and rich clusters. Blue and red points represent BCGs in poor and rich clusters, respectively. Median values of the each redshift bin are also shown separately for poor and rich clusters.}
    \label{fig:evolution}
\end{center}
\end{figure}

We plot the S\'ersic index and the effective radius of BCGs as a function of redshift in Fig. \ref{fig:evolution}. In both plots, we show BCGs in poor ($\lambda \leq 30$) and rich ($\lambda > 30$) clusters separately. To better show the trends we plotted the median values corresponding to redshifts of 0.25, 0.55, and 0.75 which are the central redshifts of the bins used in this study.

There is an increase in the S\'ersic index towards lower redshifts. However, our measurements of effective radius do not suggest any significant evolution with redshift. The trends in both parameters are consistent for poor and rich clusters as shown in Fig. \ref{fig:evolution}. The mean difference of the median values of S\'ersic index for poor and rich clusters is $ \langle \Delta n \rangle = 0.19$. Similarly, the mean difference of effective radius $ \langle \Delta log(R_e) \rangle \simeq 0.10$. These findings suggest a similar evolution for the structural parameters of BCGs both in poor and rich clusters.

In the recent study of \cite{Chu2022}, they show the similar trends for the effective radius which is suggesting the no-evolution of the BCG sizes since z=0.7. \cite{Stott2011} also found little change in size when they compared a sample of high redshift (i.e. $0.8 < z < 1.3$) BCGs, where HST imaging is available, with a local sample of BCGs at z$\sim$0.25.

Our results for S\'ersic index and effective radius evolution are in contrary to \cite{Ascaso2011}. They show similar Sersic indices for their samples at low and intermediate redshifts whereas their low redshift BCGs are almost two times larger than their counterparts at intermediate redshifts. In our study, we do not see a similar size evolution and this is consistent even we split our cluster sample into poor and rich. However, our BCGs in rich clusters have larger effective radii compared to their counterparts in poor clusters. \cite{Bai2014} showed a correlation with the stellar masses of BCGs with cluster richness and they also pointed out that the mass of BCGs increases about 1.5 times from z=0.5 to z=0. Since we do not have stellar masses for this sample of BCGs we are not able to make a direct comparison. Nevertheless, we do not see an increase of the BCG sizes in the same redshift range which could be used as an indicator for the stellar mass.

\section{Summary and Conclusions}

We present the results of the structural analysis of a BCG sample in the redshift range of $0.1 < z \leq 1.0$. This is the largest sample that is used for this kind of study. The initial galaxy cluster catalog that we employ contains 3283 clusters from the CFHTLS-W1. However, after the surface brightness profile fitting procedure, we keep only reliable results for the 1685 galaxies as we described in Sec. 4. For the procedure, we used the $r$-band images obtained within the framework of CFHTLS in the field of 72 deg$^2$. Making use of the GALFIT, a surface brightness profile fitting tool, we obtained S\'ersic indices and effective radii as the structural parameters of BCGs.

To investigate any possible environmental effect, we split the cluster sample into two sub-samples based on the host cluster richness such as poor ($\lambda \leq 30$) and rich ($\lambda > 30$). The distribution of effective radii seen in Fig. \ref{fig:histo_structural} indicates that the impact of environment on the BCG evolution can be different. As BCGs reside in the center of the potential well of galaxy clusters, cannibalism and galaxy merging might be more frequent in richer clusters. \cite{Ascaso2011} also showed the correlation of the host cluster properties with the BCG structural parameters for a BCG sample at $z \sim 0$. However, our S\'ersic index distributions for poor and rich clusters are almost identical as the K-S test suggests.

We present the relation between the mean effective surface brightness and the the effective radius which is well-known as the Kormendy relation. A comparison of our results with the literature is given in Table \ref{KormendyTable} and it can be seen that our results are in good agreement with other studies in similar redshift ranges. A detailed study of local BCGs ($z\leq0.08$) by \cite{Kluge2020} also revealed a similar slope for the Kormendy relation as $3.61\pm0.13$. \cite{Chu2022} also examined BCGs from the CFHTLS but including all the Wide fields whereas we used a cluster catalog obtained solely from the W1. Because the BCG catalog for \cite{Chu2022} had not yet been released at the time when this manuscript was submitted, it was not possible to cross-match the galaxies. However, we may expect some fraction of overlaps between the clusters, hence BCGs. Keeping that overlap in mind, it seems the results of both studies are well consistent both in the slope and in the intercept of the Kormendy relation. It is also worth noting that our BCG sample is almost two-times of the \cite{Chu2022} despite their study including all the CFHTLS-Wide imaging which covers approximately 155 $\textrm deg^{2}$.
\begin{table*}[t]
    \centering
    \caption{Coefficients of the Kormendy relation obtained in our study are compared with previous studies of different redshift ranges. Slope of the relation is denoted as $a$ and the intercept is denoted as $b$.}
    \small
    \begin{tabular}{@{}ccc|cccc@{}}
        \hline
        \multicolumn{3}{c}{This Study} & \multicolumn{4}{c}{Previous Studies}  \\
        \hline
        Redshift range & a & b & Redshift range & a & b & Reference \\
        \hline
        \multirow{2}{*}{$0.1<z<0.4$} & \multirow{2}{*}{4.005} & \multirow{2}{*}{18.955} & $0.076<z<0.394$ & 3.75 & 16.40 & \cite{Samir2020}\\
        & & & $0.15<z<0.55$ & 3.44 & - & \cite{Bildfell2008}\\
        \hline
        $0.4<z<0.7$ & 4.062 & 19.141 & $0.3<z<0.6$ & 3.346 & 18.33 & \cite{Ascaso2011}\\
        \hline
        $0.7<z<1.0$ & 3.863 & 20.222 & $0.8<z<1.3$ & 2.7 & 20.3 & \cite{Stott2011}\\
        \hline
        $0.1<z<1.0$ & 3.941 & 18.955 & $ 0.3<z<0.9$ & 3.50 & 18.01 & \cite{Bai2014}\\
        \hline
         \multirow{2}{*}{$0.1<z<1.0$} & \multirow{2}{*}{3.941} & \multirow{2}{*}{18.955} & $ 0.187<z<1.8$ & 3.33 & - & \cite{Chu2021}\\
                &       &        & $0.1<z<0.7$   & 3.34 & 18.65 & \cite{Chu2022}
        \\
        \hline
    \end{tabular}
    \label{KormendyTable}
\end{table*}

In order to see any evolutionary effect, we binned our BCGs into three redshift bins (i.e. $0.1 < z \leq 0.4$, $0.4 < z \leq 0.7$ and, $0.7 < z \leq 1.0$). The offsets in the Kormendy relation for different redshift bins are mainly due to the cosmological dimming. Besides the offsets, slopes of the individual relations are consistent with each other. In Table \ref{KormendyTable} we compare our results with previous studies where we see a general agreement for the corresponding redshift range of our study.

For the size-luminosity relation we see very similar trends for all redshift ranges except the relations are slightly offset for the highest redshift bin for poor and rich clusters. However, we should note that the less number of BCGs in rich clusters, and thus in the highest redshift bin.

The little or no evolution in the BCG sizes since $z \sim 1$ also seen in other studies \citep{Stott2011,Chu2021,Chu2022} requires further investigation as some observational studies showed an increase in size \citep{Bernardi2009,Ascaso2011,Lidman2013,Bai2014,Lavoie2016} similarly to some theoretical studies \citep{DeLucia2007,Ruszkowski2009,Naab2009}. High-resolution imaging of BCGs in different epochs could provide important clues on this controversy. Such images could be obtained within the surveys of Euclid which will be launched soon.

As pointed out in Sec.2.1, the background subtraction of CFHTLS images might have removed some of the light from the BCG outskirts. Therefore, our results should be taken into account with caution for lower redshifts (i.e. $z<0.3$). An independent reduction of the raw survey images might be useful to investigate the size evolution of BCGs at low redshifts.

\section*{Acknowledgements}

This work was supported by the TUBITAK (The Scientific and Technical Research Council of Turkey) project 117F311 through the ARDEB-1001 Program. ES acknowledges the grant provided by the Türkiye Scholarships which is a government-funded higher education scholarship program run by the Republic of Türkiye for international students. ES also acknowledges the help and valued discussions with Rasha Samir. SA acknowledges support from the Scientific Research Projects Coordination Unit of Istanbul University via project BEK-46743.

Based on observations obtained with MegaPrime/MegaCam, a joint project of CFHT and CEA/IRFU, at the Canada-France-Hawaii Telescope (CFHT) which is operated by the National Research Council (NRC) of Canada, the Institut National des Science de l'Univers of the Centre National de la Recherche Scientifique (CNRS) of France, and the University of Hawaii. This work is based in part on data products produced at Terapix available at the Canadian Astronomy Data Centre as part of the Canada-France-Hawaii Telescope Legacy Survey, a collaborative project of NRC and CNRS.


\bibliographystyle{elsarticle-harv} 
\bibliography{eman}

\begin{thebibliography}{58}
\expandafter\ifx\csname natexlab\endcsname\relax\def\natexlab#1{#1}\fi
\providecommand{\url}[1]{\texttt{#1}}
\providecommand{\href}[2]{#2}
\providecommand{\path}[1]{#1}
\providecommand{\DOIprefix}{doi:}
\providecommand{\ArXivprefix}{arXiv:}
\providecommand{\URLprefix}{URL: }
\providecommand{\Pubmedprefix}{pmid:}
\providecommand{\doi}[1]{\href{http://dx.doi.org/#1}{\path{#1}}}
\providecommand{\Pubmed}[1]{\href{pmid:#1}{\path{#1}}}
\providecommand{\bibinfo}[2]{#2}
\ifx\xfnm\relax \def\xfnm[#1]{\unskip,\space#1}\fi
\bibitem[{{Aguena} et~al.(2021){Aguena}, {Benoist}, {da Costa}, {Ogando},
  {Gschwend}, {Sampaio-Santos}, {Lima}, {Maia}, {Allam}, {Avila}, {Bacon},
  {Bertin}, {Bhargava}, {Brooks}, {Carnero Rosell}, {Carrasco Kind},
  {Carretero}, {Costanzi}, {De Vicente}, {Desai}, {Diehl}, {Doel}, {Everett},
  {Evrard}, {Ferrero}, {Fert{\'e}}, {Flaugher}, {Fosalba}, {Frieman},
  {Garc{\'\i}a-Bellido}, {Giles}, {Gruendl}, {Gutierrez}, {Hinton},
  {Hollowood}, {Honscheid}, {James}, {Jeltema}, {Kuehn}, {Kuropatkin}, {Lahav},
  {Melchior}, {Miquel}, {Morgan}, {Palmese}, {Paz-Chinch{\'o}n}, {Plazas},
  {Romer}, {Sanchez}, {Santiago}, {Schubnell}, {Serrano}, {Sevilla-Noarbe},
  {Smith}, {Soares-Santos}, {Suchyta}, {Tarle}, {To}, {Tucker} and
  {Wilkinson}}]{Aguena21}
\bibinfo{author}{{Aguena}, M.}, \bibinfo{author}{{Benoist}, C.},
  \bibinfo{author}{{da Costa}, L.N.}, \bibinfo{author}{{Ogando}, R.L.C.},
  \bibinfo{author}{{Gschwend}, J.}, \bibinfo{author}{{Sampaio-Santos}, H.B.},
  \bibinfo{author}{{Lima}, M.}, \bibinfo{author}{{Maia}, M.A.G.},
  \bibinfo{author}{{Allam}, S.}, \bibinfo{author}{{Avila}, S.},
  \bibinfo{author}{{Bacon}, D.}, \bibinfo{author}{{Bertin}, E.},
  \bibinfo{author}{{Bhargava}, S.}, \bibinfo{author}{{Brooks}, D.},
  \bibinfo{author}{{Carnero Rosell}, A.}, \bibinfo{author}{{Carrasco Kind},
  M.}, \bibinfo{author}{{Carretero}, J.}, \bibinfo{author}{{Costanzi}, M.},
  \bibinfo{author}{{De Vicente}, J.}, \bibinfo{author}{{Desai}, S.},
  \bibinfo{author}{{Diehl}, H.T.}, \bibinfo{author}{{Doel}, P.},
  \bibinfo{author}{{Everett}, S.}, \bibinfo{author}{{Evrard}, A.E.},
  \bibinfo{author}{{Ferrero}, I.}, \bibinfo{author}{{Fert{\'e}}, A.},
  \bibinfo{author}{{Flaugher}, B.}, \bibinfo{author}{{Fosalba}, P.},
  \bibinfo{author}{{Frieman}, J.}, \bibinfo{author}{{Garc{\'\i}a-Bellido}, J.},
  \bibinfo{author}{{Giles}, P.}, \bibinfo{author}{{Gruendl}, R.A.},
  \bibinfo{author}{{Gutierrez}, G.}, \bibinfo{author}{{Hinton}, S.R.},
  \bibinfo{author}{{Hollowood}, D.L.}, \bibinfo{author}{{Honscheid}, K.},
  \bibinfo{author}{{James}, D.J.}, \bibinfo{author}{{Jeltema}, T.},
  \bibinfo{author}{{Kuehn}, K.}, \bibinfo{author}{{Kuropatkin}, N.},
  \bibinfo{author}{{Lahav}, O.}, \bibinfo{author}{{Melchior}, P.},
  \bibinfo{author}{{Miquel}, R.}, \bibinfo{author}{{Morgan}, R.},
  \bibinfo{author}{{Palmese}, A.}, \bibinfo{author}{{Paz-Chinch{\'o}n}, F.},
  \bibinfo{author}{{Plazas}, A.A.}, \bibinfo{author}{{Romer}, A.K.},
  \bibinfo{author}{{Sanchez}, E.}, \bibinfo{author}{{Santiago}, B.},
  \bibinfo{author}{{Schubnell}, M.}, \bibinfo{author}{{Serrano}, S.},
  \bibinfo{author}{{Sevilla-Noarbe}, I.}, \bibinfo{author}{{Smith}, M.},
  \bibinfo{author}{{Soares-Santos}, M.}, \bibinfo{author}{{Suchyta}, E.},
  \bibinfo{author}{{Tarle}, G.}, \bibinfo{author}{{To}, C.},
  \bibinfo{author}{{Tucker}, D.L.}, \bibinfo{author}{{Wilkinson}, R.D.},
  \bibinfo{year}{2021}.
\newblock \bibinfo{title}{{The WaZP galaxy cluster sample of the dark energy
  survey year 1}}.
\newblock \bibinfo{journal}{\mnras} \bibinfo{volume}{502},
  \bibinfo{pages}{4435--4456}.
\newblock \DOIprefix\doi{10.1093/mnras/stab264},
  \href{http://arxiv.org/abs/2008.08711}{{\tt arXiv:2008.08711}}.
\bibitem[{{Aragon-Salamanca} et~al.(1998){Aragon-Salamanca}, {Baugh} and
  {Kauffmann}}]{AragonSalamanca1998}
\bibinfo{author}{{Aragon-Salamanca}, A.}, \bibinfo{author}{{Baugh}, C.M.},
  \bibinfo{author}{{Kauffmann}, G.}, \bibinfo{year}{1998}.
\newblock \bibinfo{title}{{The K-band Hubble diagram for the brightest cluster
  galaxies: a test of hierarchical galaxy formation models}}.
\newblock \bibinfo{journal}{\mnras} \bibinfo{volume}{297},
  \bibinfo{pages}{427--434}.
\newblock \DOIprefix\doi{10.1046/j.1365-8711.1998.01495.x},
  \href{http://arxiv.org/abs/astro-ph/9801277}{{\tt arXiv:astro-ph/9801277}}.
\bibitem[{{Ascaso} et~al.(2011){Ascaso}, {Aguerri}, {Varela}, {Cava},
  {Bettoni}, {Moles} and {D'Onofrio}}]{Ascaso2011}
\bibinfo{author}{{Ascaso}, B.}, \bibinfo{author}{{Aguerri}, J.A.L.},
  \bibinfo{author}{{Varela}, J.}, \bibinfo{author}{{Cava}, A.},
  \bibinfo{author}{{Bettoni}, D.}, \bibinfo{author}{{Moles}, M.},
  \bibinfo{author}{{D'Onofrio}, M.}, \bibinfo{year}{2011}.
\newblock \bibinfo{title}{{Evolution of Brightest Cluster Galaxy Structural
  Parameters in the Last \raisebox{-0.5ex}\textasciitilde6 Gyr: Feedback
  Processes Versus Merger Events}}.
\newblock \bibinfo{journal}{\apj} \bibinfo{volume}{726}, \bibinfo{pages}{69}.
\newblock \DOIprefix\doi{10.1088/0004-637X/726/2/69},
  \href{http://arxiv.org/abs/1007.3264}{{\tt arXiv:1007.3264}}.
\bibitem[{{Bai} et~al.(2014){Bai}, {Yee}, {Yan}, {Lee}, {Gilbank}, {Ellingson},
  {Barrientos}, {Gladders}, {Hsieh} and {Li}}]{Bai2014}
\bibinfo{author}{{Bai}, L.}, \bibinfo{author}{{Yee}, H.K.C.},
  \bibinfo{author}{{Yan}, R.}, \bibinfo{author}{{Lee}, E.},
  \bibinfo{author}{{Gilbank}, D.G.}, \bibinfo{author}{{Ellingson}, E.},
  \bibinfo{author}{{Barrientos}, L.F.}, \bibinfo{author}{{Gladders}, M.D.},
  \bibinfo{author}{{Hsieh}, B.C.}, \bibinfo{author}{{Li}, I.H.},
  \bibinfo{year}{2014}.
\newblock \bibinfo{title}{{The Inside-out Growth of the Most Massive Galaxies
  at 0.3 < z < 0.9}}.
\newblock \bibinfo{journal}{\apj} \bibinfo{volume}{789}, \bibinfo{pages}{134}.
\newblock \DOIprefix\doi{10.1088/0004-637X/789/2/134},
  \href{http://arxiv.org/abs/1406.4149}{{\tt arXiv:1406.4149}}.
\bibitem[{{Bellstedt} et~al.(2016){Bellstedt}, {Lidman}, {Muzzin}, {Franx},
  {Guatelli}, {Hill}, {Hoekstra}, {Kurinsky}, {Labbe}, {Marchesini}, {Marsan},
  {Safavi-Naeini}, {Sif{\'o}n}, {Stefanon}, {van de Sande}, {van Dokkum} and
  {Weigel}}]{Bellstedt2016}
\bibinfo{author}{{Bellstedt}, S.}, \bibinfo{author}{{Lidman}, C.},
  \bibinfo{author}{{Muzzin}, A.}, \bibinfo{author}{{Franx}, M.},
  \bibinfo{author}{{Guatelli}, S.}, \bibinfo{author}{{Hill}, A.R.},
  \bibinfo{author}{{Hoekstra}, H.}, \bibinfo{author}{{Kurinsky}, N.},
  \bibinfo{author}{{Labbe}, I.}, \bibinfo{author}{{Marchesini}, D.},
  \bibinfo{author}{{Marsan}, Z.C.}, \bibinfo{author}{{Safavi-Naeini}, M.},
  \bibinfo{author}{{Sif{\'o}n}, C.}, \bibinfo{author}{{Stefanon}, M.},
  \bibinfo{author}{{van de Sande}, J.}, \bibinfo{author}{{van Dokkum}, P.},
  \bibinfo{author}{{Weigel}, C.}, \bibinfo{year}{2016}.
\newblock \bibinfo{title}{{The evolution in the stellar mass of brightest
  cluster galaxies over the past 10 billion years}}.
\newblock \bibinfo{journal}{\mnras} \bibinfo{volume}{460},
  \bibinfo{pages}{2862--2874}.
\newblock \DOIprefix\doi{10.1093/mnras/stw1184},
  \href{http://arxiv.org/abs/1605.02736}{{\tt arXiv:1605.02736}}.
\bibitem[{{Bernardi}(2009)}]{Bernardi2009}
\bibinfo{author}{{Bernardi}, M.}, \bibinfo{year}{2009}.
\newblock \bibinfo{title}{{Evolution in the structural properties of early-type
  brightest cluster galaxies at small lookback time and dependence on the
  environment}}.
\newblock \bibinfo{journal}{\mnras} \bibinfo{volume}{395},
  \bibinfo{pages}{1491--1506}.
\newblock \DOIprefix\doi{10.1111/j.1365-2966.2009.14601.x},
  \href{http://arxiv.org/abs/0901.1318}{{\tt arXiv:0901.1318}}.
\bibitem[{{Bernardi} et~al.(2007){Bernardi}, {Hyde}, {Sheth}, {Miller} and
  {Nichol}}]{Bernardi2007}
\bibinfo{author}{{Bernardi}, M.}, \bibinfo{author}{{Hyde}, J.B.},
  \bibinfo{author}{{Sheth}, R.K.}, \bibinfo{author}{{Miller}, C.J.},
  \bibinfo{author}{{Nichol}, R.C.}, \bibinfo{year}{2007}.
\newblock \bibinfo{title}{{The Luminosities, Sizes, and Velocity Dispersions of
  Brightest Cluster Galaxies: Implications for Formation History}}.
\newblock \bibinfo{journal}{\aj} \bibinfo{volume}{133},
  \bibinfo{pages}{1741--1755}.
\newblock \DOIprefix\doi{10.1086/511783},
  \href{http://arxiv.org/abs/astro-ph/0607117}{{\tt arXiv:astro-ph/0607117}}.
\bibitem[{{Bertin}(2012)}]{Bertin12}
\bibinfo{author}{{Bertin}, E.}, \bibinfo{year}{2012}.
\newblock \bibinfo{title}{{Displaying Digital Deep Sky Images}}, in:
  \bibinfo{editor}{{Ballester}, P.}, \bibinfo{editor}{{Egret}, D.},
  \bibinfo{editor}{{Lorente}, N.P.F.} (Eds.), \bibinfo{booktitle}{Astronomical
  Data Analysis Software and Systems XXI}, p. \bibinfo{pages}{263}.
\bibitem[{{Bildfell} et~al.(2008){Bildfell}, {Hoekstra}, {Babul} and
  {Mahdavi}}]{Bildfell2008}
\bibinfo{author}{{Bildfell}, C.}, \bibinfo{author}{{Hoekstra}, H.},
  \bibinfo{author}{{Babul}, A.}, \bibinfo{author}{{Mahdavi}, A.},
  \bibinfo{year}{2008}.
\newblock \bibinfo{title}{{Resurrecting the red from the dead: optical
  properties of BCGs in X-ray luminous clusters}}.
\newblock \bibinfo{journal}{\mnras} \bibinfo{volume}{389},
  \bibinfo{pages}{1637--1654}.
\newblock \DOIprefix\doi{10.1111/j.1365-2966.2008.13699.x},
  \href{http://arxiv.org/abs/0802.2712}{{\tt arXiv:0802.2712}}.
\bibitem[{{Brough} et~al.(2005){Brough}, {Collins}, {Burke}, {Lynam} and
  {Mann}}]{Brough2005}
\bibinfo{author}{{Brough}, S.}, \bibinfo{author}{{Collins}, C.A.},
  \bibinfo{author}{{Burke}, D.J.}, \bibinfo{author}{{Lynam}, P.D.},
  \bibinfo{author}{{Mann}, R.G.}, \bibinfo{year}{2005}.
\newblock \bibinfo{title}{{Environmental dependence of the structure of
  brightest cluster galaxies}}.
\newblock \bibinfo{journal}{\mnras} \bibinfo{volume}{364},
  \bibinfo{pages}{1354--1362}.
\newblock \DOIprefix\doi{10.1111/j.1365-2966.2005.09679.x},
  \href{http://arxiv.org/abs/astro-ph/0510065}{{\tt arXiv:astro-ph/0510065}}.
\bibitem[{{Brough} et~al.(2008){Brough}, {Couch}, {Collins}, {Jarrett}, {Burke}
  and {Mann}}]{Brough2008}
\bibinfo{author}{{Brough}, S.}, \bibinfo{author}{{Couch}, W.J.},
  \bibinfo{author}{{Collins}, C.A.}, \bibinfo{author}{{Jarrett}, T.},
  \bibinfo{author}{{Burke}, D.J.}, \bibinfo{author}{{Mann}, R.G.},
  \bibinfo{year}{2008}.
\newblock \bibinfo{title}{{The luminosity-halo mass relation for brightest
  cluster galaxies}}.
\newblock \bibinfo{journal}{\mnras} \bibinfo{volume}{385},
  \bibinfo{pages}{L103--L107}.
\newblock \DOIprefix\doi{10.1111/j.1745-3933.2008.00442.x},
  \href{http://arxiv.org/abs/0801.1170}{{\tt arXiv:0801.1170}}.
\bibitem[{{Castignani} and {Benoist}(2016)}]{Castignani16}
\bibinfo{author}{{Castignani}, G.}, \bibinfo{author}{{Benoist}, C.},
  \bibinfo{year}{2016}.
\newblock \bibinfo{title}{{A new method to assign galaxy cluster membership
  using photometric redshifts}}.
\newblock \bibinfo{journal}{\aap} \bibinfo{volume}{595}, \bibinfo{pages}{A111}.
\newblock \DOIprefix\doi{10.1051/0004-6361/201528009},
  \href{http://arxiv.org/abs/1606.08744}{{\tt arXiv:1606.08744}}.
\bibitem[{{Chu} et~al.(2021){Chu}, {Durret} and {M{\'a}rquez}}]{Chu2021}
\bibinfo{author}{{Chu}, A.}, \bibinfo{author}{{Durret}, F.},
  \bibinfo{author}{{M{\'a}rquez}, I.}, \bibinfo{year}{2021}.
\newblock \bibinfo{title}{{Physical properties of brightest cluster galaxies up
  to redshift 1.80 based on HST data}}.
\newblock \bibinfo{journal}{\aap} \bibinfo{volume}{649}, \bibinfo{pages}{A42}.
\newblock \DOIprefix\doi{10.1051/0004-6361/202040245},
  \href{http://arxiv.org/abs/2102.01557}{{\tt arXiv:2102.01557}}.
\bibitem[{{Chu} et~al.(2022){Chu}, {Sarron}, {Durret} and
  {M{\'a}rquez}}]{Chu2022}
\bibinfo{author}{{Chu}, A.}, \bibinfo{author}{{Sarron}, F.},
  \bibinfo{author}{{Durret}, F.}, \bibinfo{author}{{M{\'a}rquez}, I.},
  \bibinfo{year}{2022}.
\newblock \bibinfo{title}{{Physical properties of more than one thousand
  brightest cluster galaxies detected in the Canada France Hawaii Telescope
  Legacy Survey}}.
\newblock \bibinfo{journal}{arXiv e-prints} ,
  \bibinfo{pages}{arXiv:2206.14209}\href{http://arxiv.org/abs/2206.14209}{{\tt
  arXiv:2206.14209}}.
\bibitem[{{Coupon} et~al.(2009){Coupon}, {Ilbert}, {Kilbinger}, {McCracken},
  {Mellier}, {Arnouts}, {Bertin}, {Hudelot}, {Schultheis}, {Le F{\`e}vre}, {Le
  Brun}, {Guzzo}, {Bardelli}, {Zucca}, {Bolzonella}, {Garilli}, {Zamorani},
  {Zanichelli}, {Tresse} and {Aussel}}]{Coupon2009}
\bibinfo{author}{{Coupon}, J.}, \bibinfo{author}{{Ilbert}, O.},
  \bibinfo{author}{{Kilbinger}, M.}, \bibinfo{author}{{McCracken}, H.J.},
  \bibinfo{author}{{Mellier}, Y.}, \bibinfo{author}{{Arnouts}, S.},
  \bibinfo{author}{{Bertin}, E.}, \bibinfo{author}{{Hudelot}, P.},
  \bibinfo{author}{{Schultheis}, M.}, \bibinfo{author}{{Le F{\`e}vre}, O.},
  \bibinfo{author}{{Le Brun}, V.}, \bibinfo{author}{{Guzzo}, L.},
  \bibinfo{author}{{Bardelli}, S.}, \bibinfo{author}{{Zucca}, E.},
  \bibinfo{author}{{Bolzonella}, M.}, \bibinfo{author}{{Garilli}, B.},
  \bibinfo{author}{{Zamorani}, G.}, \bibinfo{author}{{Zanichelli}, A.},
  \bibinfo{author}{{Tresse}, L.}, \bibinfo{author}{{Aussel}, H.},
  \bibinfo{year}{2009}.
\newblock \bibinfo{title}{{Photometric redshifts for the CFHTLS T0004 deep and
  wide fields}}.
\newblock \bibinfo{journal}{\aap} \bibinfo{volume}{500},
  \bibinfo{pages}{981--998}.
\newblock \DOIprefix\doi{10.1051/0004-6361/200811413},
  \href{http://arxiv.org/abs/0811.3326}{{\tt arXiv:0811.3326}}.
\bibitem[{{De Lucia} and {Blaizot}(2007)}]{DeLucia2007}
\bibinfo{author}{{De Lucia}, G.}, \bibinfo{author}{{Blaizot}, J.},
  \bibinfo{year}{2007}.
\newblock \bibinfo{title}{{The hierarchical formation of the brightest cluster
  galaxies}}.
\newblock \bibinfo{journal}{\mnras} \bibinfo{volume}{375},
  \bibinfo{pages}{2--14}.
\newblock \DOIprefix\doi{10.1111/j.1365-2966.2006.11287.x},
  \href{http://arxiv.org/abs/astro-ph/0606519}{{\tt arXiv:astro-ph/0606519}}.
\bibitem[{{Dressler}(1984)}]{Dressler1984}
\bibinfo{author}{{Dressler}, A.}, \bibinfo{year}{1984}.
\newblock \bibinfo{title}{{Internal kinematics of galaxies in clusters. I.
  Velocity dispersions for elliptical galaxies in Coma and Virgo.}}
\newblock \bibinfo{journal}{\apj} \bibinfo{volume}{281},
  \bibinfo{pages}{512--524}.
\newblock \DOIprefix\doi{10.1086/162124}.
\bibitem[{{Dressler} et~al.(1987){Dressler}, {Lynden-Bell}, {Burstein},
  {Davies}, {Faber}, {Terlevich} and {Wegner}}]{Dressler1987}
\bibinfo{author}{{Dressler}, A.}, \bibinfo{author}{{Lynden-Bell}, D.},
  \bibinfo{author}{{Burstein}, D.}, \bibinfo{author}{{Davies}, R.L.},
  \bibinfo{author}{{Faber}, S.M.}, \bibinfo{author}{{Terlevich}, R.},
  \bibinfo{author}{{Wegner}, G.}, \bibinfo{year}{1987}.
\newblock \bibinfo{title}{{Spectroscopy and Photometry of Elliptical Galaxies.
  I. New Distance Estimator}}.
\newblock \bibinfo{journal}{\apj} \bibinfo{volume}{313}, \bibinfo{pages}{42}.
\newblock \DOIprefix\doi{10.1086/164947}.
\bibitem[{{Dubinski}(1998)}]{Dubinski1998}
\bibinfo{author}{{Dubinski}, J.}, \bibinfo{year}{1998}.
\newblock \bibinfo{title}{{The Origin of the Brightest Cluster Galaxies}}.
\newblock \bibinfo{journal}{\apj} \bibinfo{volume}{502},
  \bibinfo{pages}{141--149}.
\newblock \DOIprefix\doi{10.1086/305901},
  \href{http://arxiv.org/abs/astro-ph/9709102}{{\tt arXiv:astro-ph/9709102}}.
\bibitem[{{Euclid Collaboration} et~al.(2019){Euclid Collaboration}, {Adam},
  {Vannier}, {Maurogordato}, {Biviano}, {Adami}, {Ascaso}, {Bellagamba},
  {Benoist}, {Cappi}, {D{\'\i}az-S{\'a}nchez}, {Durret}, {Farrens}, {Gonzalez},
  {Iovino}, {Licitra}, {Maturi}, {Mei}, {Merson}, {Munari}, {Pell{\'o}},
  {Ricci}, {Rocci}, {Roncarelli}, {Sarron}, {Amoura}, {Andreon}, {Apostolakos},
  {Arnaud}, {Bardelli}, {Bartlett}, {Baugh}, {Borgani}, {Brodwin}, {Castander},
  {Castignani}, {Cucciati}, {De Lucia}, {Dubath}, {Fosalba}, {Giocoli},
  {Hoekstra}, {Mamon}, {Melin}, {Moscardini}, {Paltani}, {Radovich},
  {Sartoris}, {Schultheis}, {Sereno}, {Weller}, {Burigana}, {Carvalho},
  {Corcione}, {Kurki-Suonio}, {Lilje}, {Sirri}, {Toledo-Moreo} and
  {Zamorani}}]{EuclidCol19}
\bibinfo{author}{{Euclid Collaboration}}, \bibinfo{author}{{Adam}, R.},
  \bibinfo{author}{{Vannier}, M.}, \bibinfo{author}{{Maurogordato}, S.},
  \bibinfo{author}{{Biviano}, A.}, \bibinfo{author}{{Adami}, C.},
  \bibinfo{author}{{Ascaso}, B.}, \bibinfo{author}{{Bellagamba}, F.},
  \bibinfo{author}{{Benoist}, C.}, \bibinfo{author}{{Cappi}, A.},
  \bibinfo{author}{{D{\'\i}az-S{\'a}nchez}, A.}, \bibinfo{author}{{Durret},
  F.}, \bibinfo{author}{{Farrens}, S.}, \bibinfo{author}{{Gonzalez}, A.H.},
  \bibinfo{author}{{Iovino}, A.}, \bibinfo{author}{{Licitra}, R.},
  \bibinfo{author}{{Maturi}, M.}, \bibinfo{author}{{Mei}, S.},
  \bibinfo{author}{{Merson}, A.}, \bibinfo{author}{{Munari}, E.},
  \bibinfo{author}{{Pell{\'o}}, R.}, \bibinfo{author}{{Ricci}, M.},
  \bibinfo{author}{{Rocci}, P.F.}, \bibinfo{author}{{Roncarelli}, M.},
  \bibinfo{author}{{Sarron}, F.}, \bibinfo{author}{{Amoura}, Y.},
  \bibinfo{author}{{Andreon}, S.}, \bibinfo{author}{{Apostolakos}, N.},
  \bibinfo{author}{{Arnaud}, M.}, \bibinfo{author}{{Bardelli}, S.},
  \bibinfo{author}{{Bartlett}, J.}, \bibinfo{author}{{Baugh}, C.M.},
  \bibinfo{author}{{Borgani}, S.}, \bibinfo{author}{{Brodwin}, M.},
  \bibinfo{author}{{Castander}, F.}, \bibinfo{author}{{Castignani}, G.},
  \bibinfo{author}{{Cucciati}, O.}, \bibinfo{author}{{De Lucia}, G.},
  \bibinfo{author}{{Dubath}, P.}, \bibinfo{author}{{Fosalba}, P.},
  \bibinfo{author}{{Giocoli}, C.}, \bibinfo{author}{{Hoekstra}, H.},
  \bibinfo{author}{{Mamon}, G.A.}, \bibinfo{author}{{Melin}, J.B.},
  \bibinfo{author}{{Moscardini}, L.}, \bibinfo{author}{{Paltani}, S.},
  \bibinfo{author}{{Radovich}, M.}, \bibinfo{author}{{Sartoris}, B.},
  \bibinfo{author}{{Schultheis}, M.}, \bibinfo{author}{{Sereno}, M.},
  \bibinfo{author}{{Weller}, J.}, \bibinfo{author}{{Burigana}, C.},
  \bibinfo{author}{{Carvalho}, C.S.}, \bibinfo{author}{{Corcione}, L.},
  \bibinfo{author}{{Kurki-Suonio}, H.}, \bibinfo{author}{{Lilje}, P.B.},
  \bibinfo{author}{{Sirri}, G.}, \bibinfo{author}{{Toledo-Moreo}, R.},
  \bibinfo{author}{{Zamorani}, G.}, \bibinfo{year}{2019}.
\newblock \bibinfo{title}{{Euclid preparation. III. Galaxy cluster detection in
  the wide photometric survey, performance and algorithm selection}}.
\newblock \bibinfo{journal}{\aap} \bibinfo{volume}{627}, \bibinfo{pages}{A23}.
\newblock \DOIprefix\doi{10.1051/0004-6361/201935088},
  \href{http://arxiv.org/abs/1906.04707}{{\tt arXiv:1906.04707}}.
\bibitem[{{Fasano} et~al.(2010){Fasano}, {Bettoni}, {Ascaso}, {Tormen},
  {Poggianti}, {Valentinuzzi}, {D'Onofrio}, {Fritz}, {Moretti}, {Omizzolo},
  {Cava}, {Moles}, {Dressler}, {Couch}, {Kj{\ae}rgaard} and
  {Varela}}]{Fasano2010}
\bibinfo{author}{{Fasano}, G.}, \bibinfo{author}{{Bettoni}, D.},
  \bibinfo{author}{{Ascaso}, B.}, \bibinfo{author}{{Tormen}, G.},
  \bibinfo{author}{{Poggianti}, B.M.}, \bibinfo{author}{{Valentinuzzi}, T.},
  \bibinfo{author}{{D'Onofrio}, M.}, \bibinfo{author}{{Fritz}, J.},
  \bibinfo{author}{{Moretti}, A.}, \bibinfo{author}{{Omizzolo}, A.},
  \bibinfo{author}{{Cava}, A.}, \bibinfo{author}{{Moles}, M.},
  \bibinfo{author}{{Dressler}, A.}, \bibinfo{author}{{Couch}, W.J.},
  \bibinfo{author}{{Kj{\ae}rgaard}, P.}, \bibinfo{author}{{Varela}, J.},
  \bibinfo{year}{2010}.
\newblock \bibinfo{title}{{The shapes of BCGs and normal ellipticals in nearby
  clusters}}.
\newblock \bibinfo{journal}{\mnras} \bibinfo{volume}{404},
  \bibinfo{pages}{1490--1504}.
\newblock \DOIprefix\doi{10.1111/j.1365-2966.2010.16361.x},
  \href{http://arxiv.org/abs/1001.2701}{{\tt arXiv:1001.2701}}.
\bibitem[{{Furnell} et~al.(2021){Furnell}, {Collins}, {Kelvin}, {Baldry},
  {James}, {Manolopoulou}, {Mann}, {Giles}, {Bermeo}, {Hilton}, {Wilkinson},
  {Romer}, {Vergara}, {Bhargava}, {Stott}, {Mayers} and {Viana}}]{Furnell2021}
\bibinfo{author}{{Furnell}, K.E.}, \bibinfo{author}{{Collins}, C.A.},
  \bibinfo{author}{{Kelvin}, L.S.}, \bibinfo{author}{{Baldry}, I.K.},
  \bibinfo{author}{{James}, P.A.}, \bibinfo{author}{{Manolopoulou}, M.},
  \bibinfo{author}{{Mann}, R.G.}, \bibinfo{author}{{Giles}, P.A.},
  \bibinfo{author}{{Bermeo}, A.}, \bibinfo{author}{{Hilton}, M.},
  \bibinfo{author}{{Wilkinson}, R.}, \bibinfo{author}{{Romer}, A.K.},
  \bibinfo{author}{{Vergara}, C.}, \bibinfo{author}{{Bhargava}, S.},
  \bibinfo{author}{{Stott}, J.P.}, \bibinfo{author}{{Mayers}, J.},
  \bibinfo{author}{{Viana}, P.}, \bibinfo{year}{2021}.
\newblock \bibinfo{title}{{The growth of intracluster light in XCS-HSC galaxy
  clusters from 0.1 < z < 0.5}}.
\newblock \bibinfo{journal}{\mnras} \bibinfo{volume}{502},
  \bibinfo{pages}{2419--2437}.
\newblock \DOIprefix\doi{10.1093/mnras/stab065},
  \href{http://arxiv.org/abs/2101.01644}{{\tt arXiv:2101.01644}}.
\bibitem[{{Graham} and {Driver}(2005)}]{GrahamDriver2005}
\bibinfo{author}{{Graham}, A.W.}, \bibinfo{author}{{Driver}, S.P.},
  \bibinfo{year}{2005}.
\newblock \bibinfo{title}{{A Concise Reference to (Projected) S{\'e}rsic
  R$^{1/n}$ Quantities, Including Concentration, Profile Slopes, Petrosian
  Indices, and Kron Magnitudes}}.
\newblock \bibinfo{journal}{\pasa} \bibinfo{volume}{22},
  \bibinfo{pages}{118--127}.
\newblock \DOIprefix\doi{10.1071/AS05001},
  \href{http://arxiv.org/abs/astro-ph/0503176}{{\tt arXiv:astro-ph/0503176}}.
\bibitem[{{Hansen} et~al.(2009){Hansen}, {Sheldon}, {Wechsler} and
  {Koester}}]{Hansen2009}
\bibinfo{author}{{Hansen}, S.M.}, \bibinfo{author}{{Sheldon}, E.S.},
  \bibinfo{author}{{Wechsler}, R.H.}, \bibinfo{author}{{Koester}, B.P.},
  \bibinfo{year}{2009}.
\newblock \bibinfo{title}{{The Galaxy Content of SDSS Clusters and Groups}}.
\newblock \bibinfo{journal}{\apj} \bibinfo{volume}{699},
  \bibinfo{pages}{1333--1353}.
\newblock \DOIprefix\doi{10.1088/0004-637X/699/2/1333},
  \href{http://arxiv.org/abs/0710.3780}{{\tt arXiv:0710.3780}}.
\bibitem[{{Hudelot} et~al.(2012){Hudelot}, {Cuillandre}, {Withington},
  {Goranova}, {McCracken}, {Magnard}, {Mellier}, {Regnault}, {Betoule},
  {Aussel}, {Kavelaars}, {Fernique}, {Bonnarel}, {Ochsenbein} and
  {Ilbert}}]{Hudelot12}
\bibinfo{author}{{Hudelot}, P.}, \bibinfo{author}{{Cuillandre}, J.C.},
  \bibinfo{author}{{Withington}, K.}, \bibinfo{author}{{Goranova}, Y.},
  \bibinfo{author}{{McCracken}, H.}, \bibinfo{author}{{Magnard}, F.},
  \bibinfo{author}{{Mellier}, Y.}, \bibinfo{author}{{Regnault}, N.},
  \bibinfo{author}{{Betoule}, M.}, \bibinfo{author}{{Aussel}, H.},
  \bibinfo{author}{{Kavelaars}, J.J.}, \bibinfo{author}{{Fernique}, P.},
  \bibinfo{author}{{Bonnarel}, F.}, \bibinfo{author}{{Ochsenbein}, F.},
  \bibinfo{author}{{Ilbert}, O.}, \bibinfo{year}{2012}.
\newblock \bibinfo{title}{{VizieR Online Data Catalog: The CFHTLS Survey (T0007
  release) (Hudelot+ 2012)}}.
\newblock \bibinfo{journal}{VizieR Online Data Catalog} ,
  \bibinfo{pages}{II/317}.
\bibitem[{{Ilbert} et~al.(2006){Ilbert}, {Arnouts}, {McCracken}, {Bolzonella},
  {Bertin}, {Le F{\`e}vre}, {Mellier}, {Zamorani}, {Pell{\`o}}, {Iovino},
  {Tresse}, {Le Brun}, {Bottini}, {Garilli}, {Maccagni}, {Picat}, {Scaramella},
  {Scodeggio}, {Vettolani}, {Zanichelli}, {Adami}, {Bardelli}, {Cappi},
  {Charlot}, {Ciliegi}, {Contini}, {Cucciati}, {Foucaud}, {Franzetti},
  {Gavignaud}, {Guzzo}, {Marano}, {Marinoni}, {Mazure}, {Meneux}, {Merighi},
  {Paltani}, {Pollo}, {Pozzetti}, {Radovich}, {Zucca}, {Bondi}, {Bongiorno},
  {Busarello}, {de La Torre}, {Gregorini}, {Lamareille}, {Mathez}, {Merluzzi},
  {Ripepi}, {Rizzo} and {Vergani}}]{Ilbert06}
\bibinfo{author}{{Ilbert}, O.}, \bibinfo{author}{{Arnouts}, S.},
  \bibinfo{author}{{McCracken}, H.J.}, \bibinfo{author}{{Bolzonella}, M.},
  \bibinfo{author}{{Bertin}, E.}, \bibinfo{author}{{Le F{\`e}vre}, O.},
  \bibinfo{author}{{Mellier}, Y.}, \bibinfo{author}{{Zamorani}, G.},
  \bibinfo{author}{{Pell{\`o}}, R.}, \bibinfo{author}{{Iovino}, A.},
  \bibinfo{author}{{Tresse}, L.}, \bibinfo{author}{{Le Brun}, V.},
  \bibinfo{author}{{Bottini}, D.}, \bibinfo{author}{{Garilli}, B.},
  \bibinfo{author}{{Maccagni}, D.}, \bibinfo{author}{{Picat}, J.P.},
  \bibinfo{author}{{Scaramella}, R.}, \bibinfo{author}{{Scodeggio}, M.},
  \bibinfo{author}{{Vettolani}, G.}, \bibinfo{author}{{Zanichelli}, A.},
  \bibinfo{author}{{Adami}, C.}, \bibinfo{author}{{Bardelli}, S.},
  \bibinfo{author}{{Cappi}, A.}, \bibinfo{author}{{Charlot}, S.},
  \bibinfo{author}{{Ciliegi}, P.}, \bibinfo{author}{{Contini}, T.},
  \bibinfo{author}{{Cucciati}, O.}, \bibinfo{author}{{Foucaud}, S.},
  \bibinfo{author}{{Franzetti}, P.}, \bibinfo{author}{{Gavignaud}, I.},
  \bibinfo{author}{{Guzzo}, L.}, \bibinfo{author}{{Marano}, B.},
  \bibinfo{author}{{Marinoni}, C.}, \bibinfo{author}{{Mazure}, A.},
  \bibinfo{author}{{Meneux}, B.}, \bibinfo{author}{{Merighi}, R.},
  \bibinfo{author}{{Paltani}, S.}, \bibinfo{author}{{Pollo}, A.},
  \bibinfo{author}{{Pozzetti}, L.}, \bibinfo{author}{{Radovich}, M.},
  \bibinfo{author}{{Zucca}, E.}, \bibinfo{author}{{Bondi}, M.},
  \bibinfo{author}{{Bongiorno}, A.}, \bibinfo{author}{{Busarello}, G.},
  \bibinfo{author}{{de La Torre}, S.}, \bibinfo{author}{{Gregorini}, L.},
  \bibinfo{author}{{Lamareille}, F.}, \bibinfo{author}{{Mathez}, G.},
  \bibinfo{author}{{Merluzzi}, P.}, \bibinfo{author}{{Ripepi}, V.},
  \bibinfo{author}{{Rizzo}, D.}, \bibinfo{author}{{Vergani}, D.},
  \bibinfo{year}{2006}.
\newblock \bibinfo{title}{{Accurate photometric redshifts for the CFHT legacy
  survey calibrated using the VIMOS VLT deep survey}}.
\newblock \bibinfo{journal}{\aap} \bibinfo{volume}{457},
  \bibinfo{pages}{841--856}.
\newblock \DOIprefix\doi{10.1051/0004-6361:20065138},
  \href{http://arxiv.org/abs/astro-ph/0603217}{{\tt arXiv:astro-ph/0603217}}.
\bibitem[{{Kluge} et~al.(2020){Kluge}, {Neureiter}, {Riffeser}, {Bender},
  {Goessl}, {Hopp}, {Schmidt}, {Ries} and {Brosch}}]{Kluge2020}
\bibinfo{author}{{Kluge}, M.}, \bibinfo{author}{{Neureiter}, B.},
  \bibinfo{author}{{Riffeser}, A.}, \bibinfo{author}{{Bender}, R.},
  \bibinfo{author}{{Goessl}, C.}, \bibinfo{author}{{Hopp}, U.},
  \bibinfo{author}{{Schmidt}, M.}, \bibinfo{author}{{Ries}, C.},
  \bibinfo{author}{{Brosch}, N.}, \bibinfo{year}{2020}.
\newblock \bibinfo{title}{{Structure of Brightest Cluster Galaxies and
  Intracluster Light}}.
\newblock \bibinfo{journal}{\apjs} \bibinfo{volume}{247}, \bibinfo{pages}{43}.
\newblock \DOIprefix\doi{10.3847/1538-4365/ab733b},
  \href{http://arxiv.org/abs/1908.08544}{{\tt arXiv:1908.08544}}.
\bibitem[{{Kormendy}(1977)}]{Kormendy1977}
\bibinfo{author}{{Kormendy}, J.}, \bibinfo{year}{1977}.
\newblock \bibinfo{title}{{Brightness distributions in compact and normal
  galaxies. II. Structure parameters of the spheroidal component.}}
\newblock \bibinfo{journal}{\apj} \bibinfo{volume}{218},
  \bibinfo{pages}{333--346}.
\newblock \DOIprefix\doi{10.1086/155687}.
\bibitem[{{Kormendy} et~al.(2009){Kormendy}, {Fisher}, {Cornell} and
  {Bender}}]{Kormendy2009}
\bibinfo{author}{{Kormendy}, J.}, \bibinfo{author}{{Fisher}, D.B.},
  \bibinfo{author}{{Cornell}, M.E.}, \bibinfo{author}{{Bender}, R.},
  \bibinfo{year}{2009}.
\newblock \bibinfo{title}{{Structure and Formation of Elliptical and Spheroidal
  Galaxies}}.
\newblock \bibinfo{journal}{\apjs} \bibinfo{volume}{182},
  \bibinfo{pages}{216--309}.
\newblock \DOIprefix\doi{10.1088/0067-0049/182/1/216},
  \href{http://arxiv.org/abs/0810.1681}{{\tt arXiv:0810.1681}}.
\bibitem[{{La Barbera} et~al.(2004){La Barbera}, {Merluzzi}, {Busarello},
  {Massarotti} and {Mercurio}}]{Barbera2004}
\bibinfo{author}{{La Barbera}, F.}, \bibinfo{author}{{Merluzzi}, P.},
  \bibinfo{author}{{Busarello}, G.}, \bibinfo{author}{{Massarotti}, M.},
  \bibinfo{author}{{Mercurio}, A.}, \bibinfo{year}{2004}.
\newblock \bibinfo{title}{{Probing galaxy evolution through the internal colour
  gradients, the Kormendy relations and the Photometric Plane of cluster
  galaxies at z {\ensuremath{\sim}} 0.2}}.
\newblock \bibinfo{journal}{\aap} \bibinfo{volume}{425},
  \bibinfo{pages}{797--812}.
\newblock \DOIprefix\doi{10.1051/0004-6361:20047157},
  \href{http://arxiv.org/abs/astro-ph/0307482}{{\tt arXiv:astro-ph/0307482}}.
\bibitem[{{Lauer} and {Postman}(1992)}]{Lauer1992}
\bibinfo{author}{{Lauer}, T.R.}, \bibinfo{author}{{Postman}, M.},
  \bibinfo{year}{1992}.
\newblock \bibinfo{title}{{The Hubble Flow from Brightest Cluster Galaxies}}.
\newblock \bibinfo{journal}{\apjl} \bibinfo{volume}{400}, \bibinfo{pages}{L47}.
\newblock \DOIprefix\doi{10.1086/186646}.
\bibitem[{{Lavoie} et~al.(2016){Lavoie}, {Willis}, {D{\'e}mocl{\`e}s},
  {Eckert}, {Gastaldello}, {Smith}, {Lidman}, {Adami}, {Pacaud}, {Pierre},
  {Clerc}, {Giles}, {Lieu}, {Chiappetti}, {Altieri}, {Ardila}, {Baldry},
  {Bongiorno}, {Desai}, {Elyiv}, {Faccioli}, {Gardner}, {Garilli}, {Groote},
  {Guennou}, {Guzzo}, {Hopkins}, {Liske}, {McGee}, {Melnyk}, {Owers},
  {Poggianti}, {Ponman}, {Scodeggio}, {Spitler} and {Tuffs}}]{Lavoie2016}
\bibinfo{author}{{Lavoie}, S.}, \bibinfo{author}{{Willis}, J.P.},
  \bibinfo{author}{{D{\'e}mocl{\`e}s}, J.}, \bibinfo{author}{{Eckert}, D.},
  \bibinfo{author}{{Gastaldello}, F.}, \bibinfo{author}{{Smith}, G.P.},
  \bibinfo{author}{{Lidman}, C.}, \bibinfo{author}{{Adami}, C.},
  \bibinfo{author}{{Pacaud}, F.}, \bibinfo{author}{{Pierre}, M.},
  \bibinfo{author}{{Clerc}, N.}, \bibinfo{author}{{Giles}, P.},
  \bibinfo{author}{{Lieu}, M.}, \bibinfo{author}{{Chiappetti}, L.},
  \bibinfo{author}{{Altieri}, B.}, \bibinfo{author}{{Ardila}, F.},
  \bibinfo{author}{{Baldry}, I.}, \bibinfo{author}{{Bongiorno}, A.},
  \bibinfo{author}{{Desai}, S.}, \bibinfo{author}{{Elyiv}, A.},
  \bibinfo{author}{{Faccioli}, L.}, \bibinfo{author}{{Gardner}, B.},
  \bibinfo{author}{{Garilli}, B.}, \bibinfo{author}{{Groote}, M.W.},
  \bibinfo{author}{{Guennou}, L.}, \bibinfo{author}{{Guzzo}, L.},
  \bibinfo{author}{{Hopkins}, A.M.}, \bibinfo{author}{{Liske}, J.},
  \bibinfo{author}{{McGee}, S.}, \bibinfo{author}{{Melnyk}, O.},
  \bibinfo{author}{{Owers}, M.S.}, \bibinfo{author}{{Poggianti}, B.},
  \bibinfo{author}{{Ponman}, T.J.}, \bibinfo{author}{{Scodeggio}, M.},
  \bibinfo{author}{{Spitler}, L.}, \bibinfo{author}{{Tuffs}, R.J.},
  \bibinfo{year}{2016}.
\newblock \bibinfo{title}{{The XXL survey XV: evidence for dry merger driven
  BCG growth in XXL-100-GC X-ray clusters}}.
\newblock \bibinfo{journal}{\mnras} \bibinfo{volume}{462},
  \bibinfo{pages}{4141--4156}.
\newblock \DOIprefix\doi{10.1093/mnras/stw1906},
  \href{http://arxiv.org/abs/1608.01223}{{\tt arXiv:1608.01223}}.
\bibitem[{{Le F{\`e}vre} et~al.(2005){Le F{\`e}vre}, {Vettolani}, {Garilli},
  {Tresse}, {Bottini}, {Le Brun}, {Maccagni}, {Picat}, {Scaramella},
  {Scodeggio}, {Zanichelli}, {Adami}, {Arnaboldi}, {Arnouts}, {Bardelli},
  {Bolzonella}, {Cappi}, {Charlot}, {Ciliegi}, {Contini}, {Foucaud},
  {Franzetti}, {Gavignaud}, {Guzzo}, {Ilbert}, {Iovino}, {McCracken}, {Marano},
  {Marinoni}, {Mathez}, {Mazure}, {Meneux}, {Merighi}, {Paltani}, {Pell{\`o}},
  {Pollo}, {Pozzetti}, {Radovich}, {Zamorani}, {Zucca}, {Bondi}, {Bongiorno},
  {Busarello}, {Lamareille}, {Mellier}, {Merluzzi}, {Ripepi} and
  {Rizzo}}]{LeFevre05}
\bibinfo{author}{{Le F{\`e}vre}, O.}, \bibinfo{author}{{Vettolani}, G.},
  \bibinfo{author}{{Garilli}, B.}, \bibinfo{author}{{Tresse}, L.},
  \bibinfo{author}{{Bottini}, D.}, \bibinfo{author}{{Le Brun}, V.},
  \bibinfo{author}{{Maccagni}, D.}, \bibinfo{author}{{Picat}, J.P.},
  \bibinfo{author}{{Scaramella}, R.}, \bibinfo{author}{{Scodeggio}, M.},
  \bibinfo{author}{{Zanichelli}, A.}, \bibinfo{author}{{Adami}, C.},
  \bibinfo{author}{{Arnaboldi}, M.}, \bibinfo{author}{{Arnouts}, S.},
  \bibinfo{author}{{Bardelli}, S.}, \bibinfo{author}{{Bolzonella}, M.},
  \bibinfo{author}{{Cappi}, A.}, \bibinfo{author}{{Charlot}, S.},
  \bibinfo{author}{{Ciliegi}, P.}, \bibinfo{author}{{Contini}, T.},
  \bibinfo{author}{{Foucaud}, S.}, \bibinfo{author}{{Franzetti}, P.},
  \bibinfo{author}{{Gavignaud}, I.}, \bibinfo{author}{{Guzzo}, L.},
  \bibinfo{author}{{Ilbert}, O.}, \bibinfo{author}{{Iovino}, A.},
  \bibinfo{author}{{McCracken}, H.J.}, \bibinfo{author}{{Marano}, B.},
  \bibinfo{author}{{Marinoni}, C.}, \bibinfo{author}{{Mathez}, G.},
  \bibinfo{author}{{Mazure}, A.}, \bibinfo{author}{{Meneux}, B.},
  \bibinfo{author}{{Merighi}, R.}, \bibinfo{author}{{Paltani}, S.},
  \bibinfo{author}{{Pell{\`o}}, R.}, \bibinfo{author}{{Pollo}, A.},
  \bibinfo{author}{{Pozzetti}, L.}, \bibinfo{author}{{Radovich}, M.},
  \bibinfo{author}{{Zamorani}, G.}, \bibinfo{author}{{Zucca}, E.},
  \bibinfo{author}{{Bondi}, M.}, \bibinfo{author}{{Bongiorno}, A.},
  \bibinfo{author}{{Busarello}, G.}, \bibinfo{author}{{Lamareille}, F.},
  \bibinfo{author}{{Mellier}, Y.}, \bibinfo{author}{{Merluzzi}, P.},
  \bibinfo{author}{{Ripepi}, V.}, \bibinfo{author}{{Rizzo}, D.},
  \bibinfo{year}{2005}.
\newblock \bibinfo{title}{{The VIMOS VLT deep survey. First epoch VVDS-deep
  survey: 11 564 spectra with 17.5 {\ensuremath{\leq}} IAB {\ensuremath{\leq}}
  24, and the redshift distribution over 0 {\ensuremath{\leq}} z
  {\ensuremath{\leq}} 5}}.
\newblock \bibinfo{journal}{\aap} \bibinfo{volume}{439},
  \bibinfo{pages}{845--862}.
\newblock \DOIprefix\doi{10.1051/0004-6361:20041960},
  \href{http://arxiv.org/abs/astro-ph/0409133}{{\tt arXiv:astro-ph/0409133}}.
\bibitem[{{Lidman} et~al.(2013){Lidman}, {Iacobuta}, {Bauer}, {Barrientos},
  {Cerulo}, {Couch}, {Delaye}, {Demarco}, {Ellingson}, {Faloon}, {Gilbank},
  {Huertas-Company}, {Mei}, {Meyers}, {Muzzin}, {Noble}, {Nantais}, {Rettura},
  {Rosati}, {S{\'a}nchez-Janssen}, {Strazzullo}, {Webb}, {Wilson}, {Yan} and
  {Yee}}]{Lidman2013}
\bibinfo{author}{{Lidman}, C.}, \bibinfo{author}{{Iacobuta}, G.},
  \bibinfo{author}{{Bauer}, A.E.}, \bibinfo{author}{{Barrientos}, L.F.},
  \bibinfo{author}{{Cerulo}, P.}, \bibinfo{author}{{Couch}, W.J.},
  \bibinfo{author}{{Delaye}, L.}, \bibinfo{author}{{Demarco}, R.},
  \bibinfo{author}{{Ellingson}, E.}, \bibinfo{author}{{Faloon}, A.J.},
  \bibinfo{author}{{Gilbank}, D.}, \bibinfo{author}{{Huertas-Company}, M.},
  \bibinfo{author}{{Mei}, S.}, \bibinfo{author}{{Meyers}, J.},
  \bibinfo{author}{{Muzzin}, A.}, \bibinfo{author}{{Noble}, A.},
  \bibinfo{author}{{Nantais}, J.}, \bibinfo{author}{{Rettura}, A.},
  \bibinfo{author}{{Rosati}, P.}, \bibinfo{author}{{S{\'a}nchez-Janssen}, R.},
  \bibinfo{author}{{Strazzullo}, V.}, \bibinfo{author}{{Webb}, T.M.A.},
  \bibinfo{author}{{Wilson}, G.}, \bibinfo{author}{{Yan}, R.},
  \bibinfo{author}{{Yee}, H.K.C.}, \bibinfo{year}{2013}.
\newblock \bibinfo{title}{{The importance of major mergers in the build up of
  stellar mass in brightest cluster galaxies at z = 1}}.
\newblock \bibinfo{journal}{\mnras} \bibinfo{volume}{433},
  \bibinfo{pages}{825--837}.
\newblock \DOIprefix\doi{10.1093/mnras/stt777},
  \href{http://arxiv.org/abs/1305.0882}{{\tt arXiv:1305.0882}}.
\bibitem[{{Lidman} et~al.(2012){Lidman}, {Suherli}, {Muzzin}, {Wilson},
  {Demarco}, {Brough}, {Rettura}, {Cox}, {DeGroot}, {Yee}, {Gilbank},
  {Hoekstra}, {Balogh}, {Ellingson}, {Hicks}, {Nantais}, {Noble}, {Lacy},
  {Surace} and {Webb}}]{Lidman2012}
\bibinfo{author}{{Lidman}, C.}, \bibinfo{author}{{Suherli}, J.},
  \bibinfo{author}{{Muzzin}, A.}, \bibinfo{author}{{Wilson}, G.},
  \bibinfo{author}{{Demarco}, R.}, \bibinfo{author}{{Brough}, S.},
  \bibinfo{author}{{Rettura}, A.}, \bibinfo{author}{{Cox}, J.},
  \bibinfo{author}{{DeGroot}, A.}, \bibinfo{author}{{Yee}, H.K.C.},
  \bibinfo{author}{{Gilbank}, D.}, \bibinfo{author}{{Hoekstra}, H.},
  \bibinfo{author}{{Balogh}, M.}, \bibinfo{author}{{Ellingson}, E.},
  \bibinfo{author}{{Hicks}, A.}, \bibinfo{author}{{Nantais}, J.},
  \bibinfo{author}{{Noble}, A.}, \bibinfo{author}{{Lacy}, M.},
  \bibinfo{author}{{Surace}, J.}, \bibinfo{author}{{Webb}, T.},
  \bibinfo{year}{2012}.
\newblock \bibinfo{title}{{Evidence for significant growth in the stellar mass
  of brightest cluster galaxies over the past 10 billion years}}.
\newblock \bibinfo{journal}{\mnras} \bibinfo{volume}{427},
  \bibinfo{pages}{550--568}.
\newblock \DOIprefix\doi{10.1111/j.1365-2966.2012.21984.x},
  \href{http://arxiv.org/abs/1208.5143}{{\tt arXiv:1208.5143}}.
\bibitem[{{Liu} et~al.(2009){Liu}, {Mao}, {Deng}, {Xia} and {Wen}}]{Liu2009}
\bibinfo{author}{{Liu}, F.S.}, \bibinfo{author}{{Mao}, S.},
  \bibinfo{author}{{Deng}, Z.G.}, \bibinfo{author}{{Xia}, X.Y.},
  \bibinfo{author}{{Wen}, Z.L.}, \bibinfo{year}{2009}.
\newblock \bibinfo{title}{{Major dry mergers in early-type brightest cluster
  galaxies}}.
\newblock \bibinfo{journal}{\mnras} \bibinfo{volume}{396},
  \bibinfo{pages}{2003--2010}.
\newblock \DOIprefix\doi{10.1111/j.1365-2966.2009.14907.x},
  \href{http://arxiv.org/abs/0904.2379}{{\tt arXiv:0904.2379}}.
\bibitem[{{Longhetti} et~al.(2007){Longhetti}, {Saracco}, {Severgnini}, {Della
  Ceca}, {Mannucci}, {Bender}, {Drory}, {Feulner} and {Hopp}}]{Longhetti2007}
\bibinfo{author}{{Longhetti}, M.}, \bibinfo{author}{{Saracco}, P.},
  \bibinfo{author}{{Severgnini}, P.}, \bibinfo{author}{{Della Ceca}, R.},
  \bibinfo{author}{{Mannucci}, F.}, \bibinfo{author}{{Bender}, R.},
  \bibinfo{author}{{Drory}, N.}, \bibinfo{author}{{Feulner}, G.},
  \bibinfo{author}{{Hopp}, U.}, \bibinfo{year}{2007}.
\newblock \bibinfo{title}{{The Kormendy relation of massive elliptical galaxies
  at z \raisebox{-0.5ex}\textasciitilde 1.5: evidence for size evolution}}.
\newblock \bibinfo{journal}{\mnras} \bibinfo{volume}{374},
  \bibinfo{pages}{614--626}.
\newblock \DOIprefix\doi{10.1111/j.1365-2966.2006.11171.x},
  \href{http://arxiv.org/abs/astro-ph/0610241}{{\tt arXiv:astro-ph/0610241}}.
\bibitem[{{Naab} et~al.(2009){Naab}, {Johansson} and {Ostriker}}]{Naab2009}
\bibinfo{author}{{Naab}, T.}, \bibinfo{author}{{Johansson}, P.H.},
  \bibinfo{author}{{Ostriker}, J.P.}, \bibinfo{year}{2009}.
\newblock \bibinfo{title}{{Minor Mergers and the Size Evolution of Elliptical
  Galaxies}}.
\newblock \bibinfo{journal}{\apjl} \bibinfo{volume}{699},
  \bibinfo{pages}{L178--L182}.
\newblock \DOIprefix\doi{10.1088/0004-637X/699/2/L178},
  \href{http://arxiv.org/abs/0903.1636}{{\tt arXiv:0903.1636}}.
\bibitem[{{Nelson} et~al.(2002){Nelson}, {Gonzalez}, {Zaritsky} and
  {Dalcanton}}]{Nelson2002}
\bibinfo{author}{{Nelson}, A.E.}, \bibinfo{author}{{Gonzalez}, A.H.},
  \bibinfo{author}{{Zaritsky}, D.}, \bibinfo{author}{{Dalcanton}, J.J.},
  \bibinfo{year}{2002}.
\newblock \bibinfo{title}{{Revisiting Brightest Cluster Galaxy Evolution with
  the Las Campanas Distant Cluster Survey}}.
\newblock \bibinfo{journal}{\apj} \bibinfo{volume}{566},
  \bibinfo{pages}{103--122}.
\newblock \DOIprefix\doi{10.1086/338054},
  \href{http://arxiv.org/abs/astro-ph/0110310}{{\tt arXiv:astro-ph/0110310}}.
\bibitem[{{Niederste-Ostholt} et~al.(2010){Niederste-Ostholt}, {Strauss},
  {Dong}, {Koester} and {McKay}}]{Niederste-Ostholt2010}
\bibinfo{author}{{Niederste-Ostholt}, M.}, \bibinfo{author}{{Strauss}, M.A.},
  \bibinfo{author}{{Dong}, F.}, \bibinfo{author}{{Koester}, B.P.},
  \bibinfo{author}{{McKay}, T.A.}, \bibinfo{year}{2010}.
\newblock \bibinfo{title}{{Alignment of brightest cluster galaxies with their
  host clusters}}.
\newblock \bibinfo{journal}{\mnras} \bibinfo{volume}{405},
  \bibinfo{pages}{2023--2036}.
\newblock \DOIprefix\doi{10.1111/j.1365-2966.2010.16597.x},
  \href{http://arxiv.org/abs/1003.0322}{{\tt arXiv:1003.0322}}.
\bibitem[{{Oegerle} and {Hill}(2001)}]{Oegerle2001}
\bibinfo{author}{{Oegerle}, W.R.}, \bibinfo{author}{{Hill}, J.M.},
  \bibinfo{year}{2001}.
\newblock \bibinfo{title}{{Dynamics of cD Clusters of Galaxies. IV. Conclusion
  of a Survey of 25 Abell Clusters}}.
\newblock \bibinfo{journal}{\aj} \bibinfo{volume}{122},
  \bibinfo{pages}{2858--2873}.
\newblock \DOIprefix\doi{10.1086/323536}.
\bibitem[{{Peng} et~al.(2002){Peng}, {Ho}, {Impey} and {Rix}}]{Peng2002}
\bibinfo{author}{{Peng}, C.Y.}, \bibinfo{author}{{Ho}, L.C.},
  \bibinfo{author}{{Impey}, C.D.}, \bibinfo{author}{{Rix}, H.W.},
  \bibinfo{year}{2002}.
\newblock \bibinfo{title}{{Detailed Structural Decomposition of Galaxy
  Images}}.
\newblock \bibinfo{journal}{\aj} \bibinfo{volume}{124},
  \bibinfo{pages}{266--293}.
\newblock \DOIprefix\doi{10.1086/340952},
  \href{http://arxiv.org/abs/astro-ph/0204182}{{\tt arXiv:astro-ph/0204182}}.
\bibitem[{{Peng} et~al.(2010){Peng}, {Ho}, {Impey} and {Rix}}]{Peng2010}
\bibinfo{author}{{Peng}, C.Y.}, \bibinfo{author}{{Ho}, L.C.},
  \bibinfo{author}{{Impey}, C.D.}, \bibinfo{author}{{Rix}, H.W.},
  \bibinfo{year}{2010}.
\newblock \bibinfo{title}{{Detailed Decomposition of Galaxy Images. II. Beyond
  Axisymmetric Models}}.
\newblock \bibinfo{journal}{\aj} \bibinfo{volume}{139},
  \bibinfo{pages}{2097--2129}.
\newblock \DOIprefix\doi{10.1088/0004-6256/139/6/2097},
  \href{http://arxiv.org/abs/0912.0731}{{\tt arXiv:0912.0731}}.
\bibitem[{{Postman} and {Lauer}(1995)}]{PostmanLauer1995}
\bibinfo{author}{{Postman}, M.}, \bibinfo{author}{{Lauer}, T.R.},
  \bibinfo{year}{1995}.
\newblock \bibinfo{title}{{Brightest Cluster Galaxies as Standard Candles}}.
\newblock \bibinfo{journal}{\apj} \bibinfo{volume}{440}, \bibinfo{pages}{28}.
\newblock \DOIprefix\doi{10.1086/175245}.
\bibitem[{{Ruszkowski} and {Springel}(2009)}]{Ruszkowski2009}
\bibinfo{author}{{Ruszkowski}, M.}, \bibinfo{author}{{Springel}, V.},
  \bibinfo{year}{2009}.
\newblock \bibinfo{title}{{The Role of Dry Mergers for the Formation and
  Evolution of Brightest Cluster Galaxies}}.
\newblock \bibinfo{journal}{\apj} \bibinfo{volume}{696},
  \bibinfo{pages}{1094--1102}.
\newblock \DOIprefix\doi{10.1088/0004-637X/696/2/1094},
  \href{http://arxiv.org/abs/0902.0373}{{\tt arXiv:0902.0373}}.
\bibitem[{{Samir} et~al.(2020){Samir}, {Takey} and {Shaker}}]{Samir2020}
\bibinfo{author}{{Samir}, R.M.}, \bibinfo{author}{{Takey}, A.},
  \bibinfo{author}{{Shaker}, A.A.}, \bibinfo{year}{2020}.
\newblock \bibinfo{title}{{The fundamental plane of brightest cluster galaxies
  and isolated elliptical galaxies}}.
\newblock \bibinfo{journal}{\apss} \bibinfo{volume}{365}, \bibinfo{pages}{142}.
\newblock \DOIprefix\doi{10.1007/s10509-020-03857-8}.
\bibitem[{{Scarlata} et~al.(2007){Scarlata}, {Carollo}, {Lilly}, {Feldmann},
  {Kampczyk}, {Renzini}, {Cimatti}, {Halliday}, {Daddi}, {Sargent},
  {Koekemoer}, {Scoville}, {Kneib}, {Leauthaud}, {Massey}, {Rhodes}, {Tasca},
  {Capak}, {McCracken}, {Mobasher}, {Taniguchi}, {Thompson}, {Ajiki}, {Aussel},
  {Murayama}, {Sanders}, {Sasaki}, {Shioya} and {Takahashi}}]{Scarlata2007}
\bibinfo{author}{{Scarlata}, C.}, \bibinfo{author}{{Carollo}, C.M.},
  \bibinfo{author}{{Lilly}, S.J.}, \bibinfo{author}{{Feldmann}, R.},
  \bibinfo{author}{{Kampczyk}, P.}, \bibinfo{author}{{Renzini}, A.},
  \bibinfo{author}{{Cimatti}, A.}, \bibinfo{author}{{Halliday}, C.},
  \bibinfo{author}{{Daddi}, E.}, \bibinfo{author}{{Sargent}, M.T.},
  \bibinfo{author}{{Koekemoer}, A.}, \bibinfo{author}{{Scoville}, N.},
  \bibinfo{author}{{Kneib}, J.P.}, \bibinfo{author}{{Leauthaud}, A.},
  \bibinfo{author}{{Massey}, R.}, \bibinfo{author}{{Rhodes}, J.},
  \bibinfo{author}{{Tasca}, L.}, \bibinfo{author}{{Capak}, P.},
  \bibinfo{author}{{McCracken}, H.J.}, \bibinfo{author}{{Mobasher}, B.},
  \bibinfo{author}{{Taniguchi}, Y.}, \bibinfo{author}{{Thompson}, D.},
  \bibinfo{author}{{Ajiki}, M.}, \bibinfo{author}{{Aussel}, H.},
  \bibinfo{author}{{Murayama}, T.}, \bibinfo{author}{{Sanders}, D.B.},
  \bibinfo{author}{{Sasaki}, S.}, \bibinfo{author}{{Shioya}, Y.},
  \bibinfo{author}{{Takahashi}, M.}, \bibinfo{year}{2007}.
\newblock \bibinfo{title}{{The Redshift Evolution of Early-Type Galaxies in
  COSMOS: Do Massive Early-Type Galaxies Form by Dry Mergers?}}
\newblock \bibinfo{journal}{\apjs} \bibinfo{volume}{172},
  \bibinfo{pages}{494--510}.
\newblock \DOIprefix\doi{10.1086/517972},
  \href{http://arxiv.org/abs/astro-ph/0701746}{{\tt arXiv:astro-ph/0701746}}.
\bibitem[{{S{\'e}rsic}(1963)}]{Sersic63}
\bibinfo{author}{{S{\'e}rsic}, J.L.}, \bibinfo{year}{1963}.
\newblock \bibinfo{title}{{Influence of the atmospheric and instrumental
  dispersion on the brightness distribution in a galaxy}}.
\newblock \bibinfo{journal}{Boletin de la Asociacion Argentina de Astronomia La
  Plata Argentina} \bibinfo{volume}{6}, \bibinfo{pages}{41--43}.
\bibitem[{{Sersic}(1968)}]{Sersic1968}
\bibinfo{author}{{Sersic}, J.L.}, \bibinfo{year}{1968}.
\newblock \bibinfo{title}{{Atlas de Galaxias Australes}}.
\bibitem[{{Stott} et~al.(2011){Stott}, {Collins}, {Burke}, {Hamilton-Morris}
  and {Smith}}]{Stott2011}
\bibinfo{author}{{Stott}, J.P.}, \bibinfo{author}{{Collins}, C.A.},
  \bibinfo{author}{{Burke}, C.}, \bibinfo{author}{{Hamilton-Morris}, V.},
  \bibinfo{author}{{Smith}, G.P.}, \bibinfo{year}{2011}.
\newblock \bibinfo{title}{{Little change in the sizes of the most massive
  galaxies since z = 1}}.
\newblock \bibinfo{journal}{\mnras} \bibinfo{volume}{414},
  \bibinfo{pages}{445--457}.
\newblock \DOIprefix\doi{10.1111/j.1365-2966.2011.18404.x},
  \href{http://arxiv.org/abs/1101.4652}{{\tt arXiv:1101.4652}}.
\bibitem[{{Stott} et~al.(2008){Stott}, {Edge}, {Smith}, {Swinbank} and
  {Ebeling}}]{Stott2008}
\bibinfo{author}{{Stott}, J.P.}, \bibinfo{author}{{Edge}, A.C.},
  \bibinfo{author}{{Smith}, G.P.}, \bibinfo{author}{{Swinbank}, A.M.},
  \bibinfo{author}{{Ebeling}, H.}, \bibinfo{year}{2008}.
\newblock \bibinfo{title}{{Near-infrared evolution of brightest cluster
  galaxies in the most X-ray luminous clusters since z = 1}}.
\newblock \bibinfo{journal}{\mnras} \bibinfo{volume}{384},
  \bibinfo{pages}{1502--1510}.
\newblock \DOIprefix\doi{10.1111/j.1365-2966.2007.12807.x},
  \href{http://arxiv.org/abs/0712.0496}{{\tt arXiv:0712.0496}}.
\bibitem[{{Tortorelli} et~al.(2018){Tortorelli}, {Mercurio}, {Paolillo},
  {Rosati}, {Gargiulo}, {Gobat}, {Balestra}, {Caminha}, {Annunziatella},
  {Grillo}, {Lombardi}, {Nonino}, {Rettura}, {Sartoris} and
  {Strazzullo}}]{Tortorelli2018}
\bibinfo{author}{{Tortorelli}, L.}, \bibinfo{author}{{Mercurio}, A.},
  \bibinfo{author}{{Paolillo}, M.}, \bibinfo{author}{{Rosati}, P.},
  \bibinfo{author}{{Gargiulo}, A.}, \bibinfo{author}{{Gobat}, R.},
  \bibinfo{author}{{Balestra}, I.}, \bibinfo{author}{{Caminha}, G.B.},
  \bibinfo{author}{{Annunziatella}, M.}, \bibinfo{author}{{Grillo}, C.},
  \bibinfo{author}{{Lombardi}, M.}, \bibinfo{author}{{Nonino}, M.},
  \bibinfo{author}{{Rettura}, A.}, \bibinfo{author}{{Sartoris}, B.},
  \bibinfo{author}{{Strazzullo}, V.}, \bibinfo{year}{2018}.
\newblock \bibinfo{title}{{The Kormendy relation of galaxies in the Frontier
  Fields clusters: Abell S1063 and MACS J1149.5+2223}}.
\newblock \bibinfo{journal}{\mnras} \bibinfo{volume}{477},
  \bibinfo{pages}{648--668}.
\newblock \DOIprefix\doi{10.1093/mnras/sty617},
  \href{http://arxiv.org/abs/1803.02375}{{\tt arXiv:1803.02375}}.
\bibitem[{{Ulgen} et~al.(2022){Ulgen}, {Alis}, {Benoist}, {Yelkenci}, {Cakir},
  {Fisek} and {Karatas}}]{Ulgen2022}
\bibinfo{author}{{Ulgen}, E.K.}, \bibinfo{author}{{Alis}, S.},
  \bibinfo{author}{{Benoist}, C.}, \bibinfo{author}{{Yelkenci}, F.K.},
  \bibinfo{author}{{Cakir}, O.}, \bibinfo{author}{{Fisek}, S.},
  \bibinfo{author}{{Karatas}, Y.}, \bibinfo{year}{2022}.
\newblock \bibinfo{title}{{Identification and properties of isolated field
  elliptical galaxies from CFHTLS-W1}}.
\newblock \bibinfo{journal}{\pasa} \bibinfo{volume}{39}, \bibinfo{pages}{e031}.
\newblock \DOIprefix\doi{10.1017/pasa.2022.28},
  \href{http://arxiv.org/abs/2205.10669}{{\tt arXiv:2205.10669}}.
\bibitem[{{Vikhlinin} et~al.(2006){Vikhlinin}, {Kravtsov}, {Forman}, {Jones},
  {Markevitch}, {Murray} and {Van Speybroeck}}]{Vikhlinin2006}
\bibinfo{author}{{Vikhlinin}, A.}, \bibinfo{author}{{Kravtsov}, A.},
  \bibinfo{author}{{Forman}, W.}, \bibinfo{author}{{Jones}, C.},
  \bibinfo{author}{{Markevitch}, M.}, \bibinfo{author}{{Murray}, S.S.},
  \bibinfo{author}{{Van Speybroeck}, L.}, \bibinfo{year}{2006}.
\newblock \bibinfo{title}{{Chandra Sample of Nearby Relaxed Galaxy Clusters:
  Mass, Gas Fraction, and Mass-Temperature Relation}}.
\newblock \bibinfo{journal}{\apj} \bibinfo{volume}{640},
  \bibinfo{pages}{691--709}.
\newblock \DOIprefix\doi{10.1086/500288},
  \href{http://arxiv.org/abs/astro-ph/0507092}{{\tt arXiv:astro-ph/0507092}}.
\bibitem[{{Von Der Linden} et~al.(2007){Von Der Linden}, {Best}, {Kauffmann}
  and {White}}]{VonDerLinden2007}
\bibinfo{author}{{Von Der Linden}, A.}, \bibinfo{author}{{Best}, P.N.},
  \bibinfo{author}{{Kauffmann}, G.}, \bibinfo{author}{{White}, S.D.M.},
  \bibinfo{year}{2007}.
\newblock \bibinfo{title}{{How special are brightest group and cluster
  galaxies?}}
\newblock \bibinfo{journal}{\mnras} \bibinfo{volume}{379},
  \bibinfo{pages}{867--893}.
\newblock \DOIprefix\doi{10.1111/j.1365-2966.2007.11940.x},
  \href{http://arxiv.org/abs/astro-ph/0611196}{{\tt arXiv:astro-ph/0611196}}.
\bibitem[{{West} et~al.(2017){West}, {de Propris}, {Bremer} and
  {Phillipps}}]{West2017}
\bibinfo{author}{{West}, M.J.}, \bibinfo{author}{{de Propris}, R.},
  \bibinfo{author}{{Bremer}, M.N.}, \bibinfo{author}{{Phillipps}, S.},
  \bibinfo{year}{2017}.
\newblock \bibinfo{title}{{Ten billion years of brightest cluster galaxy
  alignments}}.
\newblock \bibinfo{journal}{Nature Astronomy} \bibinfo{volume}{1},
  \bibinfo{pages}{0157}.
\newblock \DOIprefix\doi{10.1038/s41550-017-0157},
  \href{http://arxiv.org/abs/1706.03798}{{\tt arXiv:1706.03798}}.
\bibitem[{{Yelkenci}(2015)}]{Yelkenci2015}
\bibinfo{author}{{Yelkenci}, F.K.}, \bibinfo{year}{2015}.
\newblock \bibinfo{title}{{Evolution of the galaxy morphology - density
  relation;}}.
\newblock Ph.D. thesis. Istanbul University, Turkey.
\bibitem[{{Zhao} et~al.(2015){Zhao}, {Arag{\'o}n-Salamanca} and
  {Conselice}}]{Zhao2015}
\bibinfo{author}{{Zhao}, D.}, \bibinfo{author}{{Arag{\'o}n-Salamanca}, A.},
  \bibinfo{author}{{Conselice}, C.J.}, \bibinfo{year}{2015}.
\newblock \bibinfo{title}{{Evolution of the brightest cluster galaxies: the
  influence of morphology, stellar mass and environment}}.
\newblock \bibinfo{journal}{\mnras} \bibinfo{volume}{453},
  \bibinfo{pages}{4444--4455}.
\newblock \DOIprefix\doi{10.1093/mnras/stv1940},
  \href{http://arxiv.org/abs/1508.04845}{{\tt arXiv:1508.04845}}.

\end{thebibliography}

\end{document}